\documentclass[%
 reprint,
preprintnumbers,
nofootinbib,
nobibnotes,
 amsmath,amssymb,
 aps,
 prd,
]{revtex4-2}

\usepackage[utf8]{inputenc}
\usepackage{amssymb}
\usepackage{physics}
\usepackage{graphicx}
\usepackage{hyperref}
\usepackage{slashed}
\usepackage{bbm}
\usepackage{pifont}
\usepackage{xspace}
\usepackage{enumitem}
\usepackage[section]{placeins}
\usepackage{MnSymbol}

\newcommand{\be}{\begin{equation}}
\newcommand{\ee}{\end{equation}}
\newcommand{\bea}{\begin{eqnarray}}
\newcommand{\eea}{\end{eqnarray}}

\usepackage{color}

\usepackage[normalem]{ulem}

\usepackage{graphicx}
\usepackage{dcolumn}
\usepackage{bm}

\begin{document}

\title{From quantum gravity to gravitational waves through cosmic strings}

\preprint{MS-TP-23-27}

\author{Astrid Eichhorn$^1$}
\email{eichhorn@cp3.sdu.dk}
\author{Rafael R. Lino dos Santos$^{1,2}$}
 \email{rado@cp3.sdu.dk}
\author{João Lucas Miqueleto$^{1,3}$}     		 
\affiliation{%
 ${}^1\,$ CP3-Origins,  University  of  Southern  Denmark,  Campusvej  55,  DK-5230  Odense  M,  Denmark\\
 ${}^2\,$University of Münster, Institute for Theoretical Physics, 48149 Münster, Germany\\
 ${}^3\,$Centro de Ciências Naturais e Humanas, Universidade Federal do ABC, Avenida dos Estados, 5001, Santo André, São Paulo 09210-580, Brazil}%

\begin{abstract}
New physics beyond the Standard Model can give rise to stochastic gravitational-wave backgrounds, for example, through cosmic strings. In this way, gravitational-wave searches with pulsar-timing arrays as well as existing and future laser interferometers may provide information on particle physics beyond the Standard Model.

Here, we take one additional step and link particle physics beyond the Standard Model to quantum gravity. We investigate whether particle physics models that may give rise to cosmic strings can be embedded into an asymptotically safe theory of quantum gravity and matter. We focus on models where cosmic strings  arise from U(1)-symmetry breaking in an extended Yukawa-Abelian-Higgs sector that may be part of a dark sector. 
We find a negative answer for the simplest model that can give rise to cosmic strings and also find constraints on an extended model.

We tentatively conclude that cosmic strings are difficult to accommodate in asymptotically safe models. This fits well with the latest 15-year dataset and search for new physics from the NANOGrav Collaboration, which disfavors a stable-cosmic-string interpretation. In that sense, the recent data provide an indirect, albeit at present rather tentative, hint about the quantum theory of gravity.
\end{abstract}

\maketitle

\section{Motivation}
\label{sec:intro}
Gravitational waves (GWs) have become a powerful probe of fundamental physics. Besides increasingly stringent tests of General Relativity (GR) \cite{LIGOScientific:2016lio,LIGOScientific:2018dkp,LIGOScientific:2019fpa,LIGOScientific:2020tif,LIGOScientific:2021sio}, GW-observatories can provide information about cosmology and particle physics beyond the Standard Model (BSM) \cite{Weir:2017wfa,Caprini:2019egz,Auclair:2019wcv,Barausse:2020rsu,Santos:2022hlx,Caldwell:2022qsj},
and even quantum gravity \cite{Addazi:2021xuf}. After the success of the LIGO-Virgo Collaboration in detecting transient GW signals from black-hole \cite{LIGOScientific:2016aoc} and neutron-star binary mergers \cite{LIGOScientific:2017vwq}, the LIGO-Virgo-KAGRA  and Pulsar Timing Array (PTA) Collaborations are looking for \textit{stochastic gravitational waves }\cite{Caprini:2018mtu,Christensen:2018iqi}. A detection of such stochastic background hinges on finding evidence for quadrupolar Hellings-Downs correlations predicted by GR \cite{1983ApJ...265L..39H}. In the past few years, PTA searches collected positive evidence for a common red-noise process in the nHz frequency range, first announced by the North American Nanohertz Observatory for Gravitational Waves (NANOGrav) Collaboration in 2020 in their 12.5-year dataset \cite{NANOGrav:2020bcs} and followed by similar findings by the Parkes Pulsar Timing Array (PPTA) \cite{Goncharov:2021oub}, the European Pulsar Timing Array (EPTA) \cite{Chalumeau:2021fpz}, and the International Pulsar Timing Array (IPTA) \cite{Antoniadis:2022pcn} collaborations. Very recently, first evidence for a Hellings-Downs pattern was discovered in the NANOGrav dataset \cite{NANOGrav:2023gor, NANOGrav:2023pdq,NANOGrav:2023icp,NANOGrav:2023ctt,NANOGrav:2023hfp,NANOGrav:2023hvm,NANOGrav:2023hde,NANOGrav:2023tcn}, the EPTA dataset \cite{Antoniadis:2023ott,Antoniadis:2023lym,Antoniadis:2023puu,Antoniadis:2023xlr,Antoniadis:2023aac,Smarra:2023ljf}, the PPTA dataset \cite{Reardon:2023gzh,Zic:2023gta,Reardon:2023zen} and the Chinese Pulsar Timing Collaboration (CPTA) dataset \cite{Xu:2023wog}, suggesting a GW origin of the signal.

If the detection is indeed of a background of stochastic GWs, these collaborations may either have detected supermassive binary black-hole mergers or may probe physics at early times in the history of the universe \cite{NANOGrav:2023hfp,NANOGrav:2023hvm,Antoniadis:2023xlr}. These early times are not accessible with the cosmic microwave background (CMB), since the universe was transparent to the gravitational-wave background before recombination, but not to photons. At these early times, particle physics beyond the Standard Model (SM) may generate a stochastic GW background. Various models are therefore being explored, see, for instance, the NANOGrav search for signals of new physics in their latest 15-year dataset {\cite{NANOGrav:2023hvm}}. Two of these possible sources are first-order phase transitions and cosmic strings.

Within the SM, there are no first-order phase transitions and cosmic strings do not arise \cite{Kajantie:1996mn}; thus BSM physics is required. Examples for such BSM models are extensions of the Higgs sector of the SM, see, e.g., \cite{Kamionkowski:1993fg,Grojean:2006bp,Caprini:2007xq,No:2011fi,Kakizaki:2015wua,Vaskonen:2016yiu,Beniwal:2017eik,Marzola:2017jzl,Chala:2018ari,Alves:2018jsw}, or models of the dark matter \cite{Schwaller:2015tja,Breitbach:2018ddu,Bertone:2019irm,Nakai:2020oit,Ratzinger:2020koh,Morgante:2022zvc,Bringmann:2023opz}. Often, these models are introduced on an ad-hoc basis, as bottom-up approaches to new physics. Here, we take a different approach and combine bottom-up with top-down considerations: we consider the simplest models that give rise to a stochastic GW background from cosmic strings and explore whether these models make sense from a top-down point of view. Specifically, we investigate whether they are ultraviolet (UV) complete when coupled to quantum gravity.

We thereby establish a link between the properties of gravity in the UV and observations of gravity in the deep IR, in the form of a stochastic GW background. This link uses that quantum gravity constrains the particle-physics models that can give rise to a stochastic GW background. Therefore, a detection of a stochastic background, with a spectrum that is explained by a particular (class of) BSM physics, provides an indirect hint about quantum gravity: if a given theory of quantum gravity is not compatible with the particular (class of) BSM physics, such a detection would disfavor the theory.

In view of the difficulty of probing quantum gravity observationally, any hints that can be found for or against a given theory, are highly valuable. GWs (from transient events) have already been used to place constraints, e.g., on the mass of the graviton \cite{Baker:2017hug} and on Lorentz-invariance violation in gravity \cite{EmirGumrukcuoglu:2017cfa}. If found, either of the two would have profound implications for the quantization of gravity. Here, we explore a different probe of quantum gravity, namely by testing whether it can accommodate particle physics models that give rise to cosmic strings and thus produce a stochastic GW background.

This paper is structured as follows. In Sec.~\ref{sec:intro2} we review how cosmic strings arise and how the current data from pulsar timing arrays constrains cosmic strings. Further, we review the current status of asymptotically safe gravity-matter models. In Sec.~\ref{sec:AbelianHiggs} we attempt to bring cosmic strings into the asymptotic-safety paradigm, by exploring whether the Abelian Higgs model can be accommodated in the paradigm. We find a negative answer and thus extend the model in Sec.~\ref{sec:extAbelianHiggs}, where we find that the negative answer can be circumvented (under the approximations we work with) only if a relatively large number of auxiliary degrees of freedom are added to the setting. We thus conclude in Sec.~\ref{sec:conclusions} that asymptotic safety disfavors cosmic strings (although they are not ruled out). Additional technical details are provided in the appendixes.

\section{Introduction: Cosmic strings and asymptotic safety}\label{sec:intro2}
\subsection{Cosmic strings}
In quantum field theories, cosmic strings are one-dimensional topological defects that can arise in an Abelian gauge theory, when the symmetry is broken spontaneously \cite{Kibble:1976sj,Vilenkin:1984ib,Hindmarsh:1994re}. The order-parameter field takes values within a U(1) manifold in the symmetry-broken phase. To respect causality, the value of the order-parameter field that it assumes after the phase transition in widely separated regions is uncorrelated \cite{Kibble:1980mv,Kibble:1981gv}. Once the system has settled down, topologically stable line defects remain, along which the order parameter vanishes; these are the cosmic strings. 

Electroweak symmetry breaking does not result in cosmic strings because the vacuum manifold does not give rise to topologically stable line defects; thus, cosmic strings are a signature of new physics beyond the SM. Grand unified theories (GUTs) are one example of a setting in which cosmic strings may arise \cite{Kibble:1982ae}; dark gauge groups another \cite{Hyde:2013fia,Long:2014mxa}.\footnote{Dark photons may be produced by the decay of the corresponding cosmic strings and form cold dark matter today \cite{Long:2019lwl,Kitajima:2022lre}.}

Cosmic strings form a network that emits GWs during its evolution when string loops form, which oscillate due to the string tension, and when local features such as cusps or kinks emit bursts of GWs \cite{Vilenkin:1981bx,Vachaspati:1984gt,Caldwell:1991jj}. Taken together, they produce a stochastic gravitational-wave background. A first estimate for the spectrum is based on modelling the strings as Nambu-Goto strings.\footnote{Going beyond the Nambu-Goto paradigm, field-theoretic strings can also access other decay channels, e.g., into dark matter, and the dominant decay channel depends on the initial conditions \cite{Hindmarsh:2022awe}. Whether field-theoretical cosmic strings can produce gravitational-wave spectra that agree with Nambu-Goto strings simulations is subject to an ongoing debate in the literature \cite{Auclair:2019wcv,Blanco-Pillado:2023sap}.} In more detail, the GW spectrum depends on many aspects of the cosmic string network and its evolution (see, for instance, \cite{Kibble:1984hp,Ringeval:2005kr,Lorenz:2010sm,Blanco-Pillado:2011egf,Blanco-Pillado:2013qja,Blanco-Pillado:2015ana,Blanco-Pillado:2017oxo,Auclair:2019wcv,Blanco-Pillado:2023sap}), many of which are only accessible via numerical simulations. The resulting spectrum has a maximum amplitude that increases with the string tension $G\mu$. This maximum happens at a peak frequency that decreases with $G\mu$. Above this frequency, the spectrum has a very slow fall-off toward large frequencies.

GW data from PTA searches are therefore especially useful to constrain cosmic strings with large string tensions. The announcement of evidence for a common-red noise signal (albeit without evidence for the Hellings-Downs correlations that characterizes GWs) by the NANOGrav Collaboration in 2020 \cite{NANOGrav:2020bcs} triggered speculation about cosmic strings as one possible source for that background \cite{Blasi:2020mfx,Ellis:2020ena,Buchmuller:2020lbh,Samanta:2020cdk,Bian:2020urb,Blanco-Pillado:2021ygr,Ahmed:2022rwy,Afzal:2022vjx}, complemented by searches with PPTA, IPTA, and EPTA datasets \cite{Chen:2022azo,Bian:2022tju,Wang:2022rjz,EPTA:2023hof}. Although a GW spectrum sourced by \textit{stable cosmic strings} could fit the NANOGrav 12.5yr data relatively well \cite{Blasi:2020mfx,Ellis:2020ena}, later datasets \cite{Goncharov:2021oub,Chalumeau:2021fpz,Antoniadis:2022pcn} indicated that GW data actually prefer a steeper GW spectrum, while the previous stable-cosmic-string interpretation yields a too flat spectrum. As a large $G\mu$ value produces a flat GW spectrum in the PTA bandwidth, PTA searches put upper bounds on the string tension.

Very recently, the NANOGrav Collaboration performed a complete Bayesian search for cosmic strings \cite{NANOGrav:2023hvm} and set new upper bounds on the string tension for four different models of stable cosmic strings, following the numerical-simulation description of \cite{Blanco-Pillado:2015ana}. The bounds lie between $\log_{10} (G\mu) < -9.67$ and $\log_{10} (G\mu)<-10.10$  depending on the cosmic-string model. Crucially, a stable-cosmic-string interpretation of the data is disfavored, compared to a signal from a population of supermassive black-hole binaries \cite{NANOGrav:2023hvm}.\footnote{Additionally, the EPTA Collaboration also reported \cite{Antoniadis:2023xlr} an upper bound on the string tension of $\log_{10} (G\mu) < -9.77$ for a cosmic-string model following \cite{Blanco-Pillado:2015ana}. Additionally, EPTA also reported an upper bound of $\log_{10} (G\mu) < -10.44$ for the cosmic-string model by \cite{Lorenz:2010sm}. PPTA Collaboration's previous dataset \cite{Chen:2022azo} put an upper bound of $\log_{10} (G\mu) \lesssim -9.3$, following \cite{Blanco-Pillado:2015ana}. Bounds from the CMB are weaker, $G\mu \lesssim  10^{-7}$ \cite{Planck:2015fie,Charnock:2016nzm,Lizarraga:2016onn}, even though model independent. While bounds from the runs of the LIGO-Virgo-KAGRA Collaboration can be competitive  \cite{LIGOScientific:2021nrg}, they rely on a large extrapolation over frequency space. New bounds might come with future interferometric observatories such as LISA \cite{Auclair:2019wcv}, the Einstein Telescope \cite{Maggiore:2019uih}, the Cosmic Explorer, DECIGO, and BBO.}

The string tension $G\mu$ is related to the spontaneous symmetry-breaking scale $v$ through \cite{Hindmarsh:1994re}
\begin{equation}
v \sim 10^{16}\text{GeV} \left(\frac{G\mu}{10^{-7}}\right)^{1/2}. \label{eq:vev}
\end{equation}
Thus, the latest NANOGrav data constrain the scale of symmetry breaking to be $ v \lesssim 4.6 \times 10^{14}$ GeV.

Beyond the stable-cosmic-string paradigm, for which $G\mu$ is  the only parameter, there are other cosmic-string descriptions with extra parameters, such as metastable cosmic strings \cite{Buchmuller:2020lbh},  which can fit  the data better \cite{NANOGrav:2023hvm}.
 
 Finally, cosmic strings can arise in a string-theoretic context, when the cosmological expansion during inflation stretches fundamental strings so that they become classical objects \cite{Sarangi:2002yt,Jones:2003da,Dvali:2003zj,Copeland:2003bj}, see \cite{Copeland:2011dx} for a review. The resulting network of cosmic superstrings may produce GWs similar to field-theoretic strings, fitting the data better \cite{NANOGrav:2023hvm} . Within a ten-dimensional superstring-theory setting, the associated string tension is set by the corresponding fundamental Planck scale in ten dimensions. When the six extra dimensions are compactified, the tension can be reduced, such that a stochastic GW background from string theory may be compatible with the observational constraints on the string tension.

In the present paper, we investigate for the first time, whether a different candidate theory of quantum gravity, namely asymptotically safe quantum gravity, can also accommodate cosmic strings. Because asymptotically safe gravity is a quantum field theory, the corresponding strings would be field-theoretic strings arising from a phase transition. We connect our work with the latest phenomenological searches by setting the symmetry-breaking scale according to the most recent results for stable cosmic strings. 

\subsection{Asymptotically safe gravity with matter: a new landscape}
Asymptotic safety is an appealing approach to quantum gravity because of its conceptual and technical simplicity: instead of introducing new degrees of freedom, new structures, and a new mathematical framework, as many approaches do, it is based on just the metric and standard quantum field theory. The single new ingredient required to build a predictive quantum field theory of the metric is scale symmetry. This symmetry does not arise at the classical level, but is instead a genuine consequence of the quantum nature of gravity, because it is realized in the form of asymptotic safety. Asymptotic safety is quantum scale symmetry: quantum fluctuations balance out in such a way that the running, scale-dependent couplings of the theory are constant. An asymptotically safe regime can thus be discovered by searching for zeros of the beta functions of the couplings, which encode the scale dependence of couplings. Using functional Renormalization Group techniques \cite{Wetterich:1992yh,Morris:1993qb,Reuter:1996cp}, see \cite{Dupuis:2020fhh} for a recent review and Appendix~\ref{app:FRG}, compelling evidence for asymptotic safety in gravity has been found, see \cite{Reuter:2012id,Pereira:2019dbn,Bonanno:2020bil,Reichert:2020mja,Pawlowski:2020qer,Saueressig:2023irs,Percacci:2023rbo} for reviews.

Beyond pure gravity, there is strong evidence for asymptotic safety in gravity with matter\footnote{By matter we refer to all non-gravitational fields, such that, e.g., the SM gauge fields are also ``matter" in our nomenclature.}, see \cite{Eichhorn:2018yfc,Eichhorn:2022jqj,Eichhorn:2022gku} for reviews. Under the impact of quantum gravity fluctuations, the scale-dependence of matter couplings changes.  For our purposes, the following important results will be central:

\begin{enumerate}[label=(\roman*)]

\item There is strong evidence that quantum gravity fluctuations antiscreen all gauge couplings above the Planck scale \cite{Daum:2009dn,Folkerts:2011jz,Christiansen:2017cxa,Eichhorn:2021qet}, which can solve the Landau pole/triviality problem for the Abelian gauge coupling \cite{Harst:2011zx,Christiansen:2017gtg,Eichhorn:2017lry,Eichhorn:2019yzm}.\footnote{Seemingly different results are found in calculations based on perturbation theory, see, e.g., \cite{Robinson:2005fj,Pietrykowski:2006xy,Toms:2007sk,Anber:2010uj,Ellis:2010rw}, where a gravitational contribution to the beta function is or is not present, depending on the choice of scheme. In \cite{deBrito:2022vbr}, a possible reason for this difference is found: In \cite{Robinson:2005fj,Pietrykowski:2006xy,Toms:2007sk,Anber:2010uj,Ellis:2010rw}, the gravitational coupling is treated as a fixed external parameter. This may not be the correct treatment in some schemes, where the gravitational coupling may assume a fixed-point value that is divergent. In \cite{deBrito:2022vbr}, a family of schemes is analyzed, containing a case with divergent fixed-point value as a limiting case. It is shown that if the limit is taken carefully, unphysical quantities (such as the gravitational coupling or the gravitational contribution to the beta function of the gauge coupling, none of which corresponds to a measurable quantity) diverge, but a physical and universal quantity, namely the critical exponent at the fixed point, stays finite. Generalizing from this example, one may expect that perturbative calculations need to include the beta function for the gravitational coupling and focus on universal quantities (such as critical exponents), instead of unphysical quantities (such as contributions to beta functions), in order to find agreement between different schemes.} This solution restricts the low-energy value of the Abelian gauge coupling, because the requirement of scale symmetry bounds the Planck-scale value of the gauge coupling from above. Thus, a phenomenologically relevant constraint on low-energy physics follows from demanding asymptotic safety for the Abelian gauge sector with gravity.

\item Oppositely to the Abelian gauge coupling, the Higgs quartic coupling is screened by gravity fluctuations \cite{Narain:2009fy,Eichhorn:2017als,Pawlowski:2018ixd,Wetterich:2019rsn,Eichhorn:2020sbo,Pastor-Gutierrez:2022nki}, which results in a vanishing value of the Higgs quartic coupling at the Planck scale. As pointed out prior to the discovery of the Higgs particle at the LHC, asymptotic safety can thereby predict a Higgs mass close to the thereafter discovered value \cite{Shaposhnikov:2009pv}.\footnote{The value is subject to precise knowledge of the top quark mass \cite{Bezrukov:2014ina}, which is not sufficiently well known to decide whether or not the asymptotically safe prediction of the Higgs mass matches observations.}

\item Finally, asymptotically safe gravity can both antiscreen or screen Yukawa couplings, depending on the values of gravitational couplings at microscopic scales \cite{Oda:2015sma,Eichhorn:2016esv,Eichhorn:2017eht}.

\begin{enumerate}
\item In the case of screening, the SM cannot be accommodated in an asymptotically safe setting because the Yukawa couplings must vanish at the Planck scale, which translates into vanishing Yukawa couplings at the electroweak scale. Thus, compatibility with the SM constrains the values of the gravitational couplings at microscopic scales.

\item In the case of antiscreening, the value of the top quark Yukawa coupling is bounded from above  \cite{Eichhorn:2017ylw,Eichhorn:2018whv,Alkofer:2020vtb}, just like the Abelian gauge coupling. Interestingly, there are indications that the gravitational couplings satisfy the conditions required to generate antiscreening, at least if the impact of the SM matter fields on the gravitational couplings is accounted for \cite{Dona:2013qba,Eichhorn:2017ylw}, see also \cite{Pastor-Gutierrez:2022nki} for an alternative mechanism.\footnote{There is a significant systematic uncertainty associated to the values of the gravitational couplings at microscopic scales, see \cite{Eichhorn:2022gku} and references therein.}
\end{enumerate}
 
\end{enumerate}

Beyond the SM, gravity screens or antiscreens various interactions, such that an asymptotically safe landscape of models emerges. There are indications that this ``new" landscape is rather distinct from the landscape associated to string theory. First, there are indications that the asymptotically safe landscape is small, in the sense that, e.g., popular candidates for dark matter are not included in it or are strongly constrained \cite{Eichhorn:2017als,Reichert:2019car,Eichhorn:2020sbo,Eichhorn:2020kca,Boos:2022pyq}. Second, there are indications that this landscape differs when it comes to axion-like particles, which are rather prevalent in string theory \cite{Ringwald:2012cu}, but may not be compatible with asymptotic safety \cite{deBrito:2021akp}. 

Furthermore, asymptotically safe solutions to the muon- (g-2)-problem \cite{Kowalska:2020zve} and neutrino mass generation \cite{deBrito:2019epw,Domenech:2020yjf,Kowalska:2022ypk,Eichhorn:2022vgp} exist, flavor anomalies have also been explored \cite{Kowalska:2020gie,Chikkaballi:2022urc}, and dark-energy \cite{Rubio:2017gty,Eichhorn:2022ngh,Wetterich:2022brb} and inflationary models can be constrained, see, e.g.,  \cite{Bonanno:2015fga,Bonanno:2018gck, Wetterich:2019rsn,Platania:2019qvo,Eichhorn:2020kca,Chojnacki:2021fag,Sen:2022xlp}. Here, we extend such studies to investigate whether simple models that can in principle give rise to cosmic strings can be accommodated in the asymptotically safe landscape or not.
\\

Asymptotic safety is sometimes referred to as a non-perturbative UV completion. However, there are no indications that an asymptotically safe fixed point in the SM together with gravity is actually nonperturbative. Instead, the fixed point appears to be \emph{near-perturbative}, although not an asymptotically free fixed point (and in that sense not strictly perturbative). What this term means more specifically is clarified through the following list of characteristics that have been found:
\begin{enumerate}[label=(\roman*)]

\item A fixed point in the gravitational sector can be found using perturbative techniques \cite{Niedermaier:2009zz,Niedermaier:2010zz}.\footnote{These studies have not yet been extended to account for matter fields.} 

\item A fixed point in the gravitational sector shows scaling exponents which are close to the canonical dimensions of the couplings \cite{Falls:2013bv,Falls:2014tra,Falls:2017lst,Falls:2018ylp}.\footnote{For an asymptotically free fixed point, scaling exponents are exactly the canonical dimensions.} 

\item Extrapolating the SM from the electroweak scale to the Planck scale and beyond is possible within perturbation theory, with one-loop beta functions describing the running couplings reasonably well, see \cite{Alkofer:2020vtb} for an explicit study of 2-loop effects.

\item The size of the gravitational contributions to the beta functions of matter couplings is typically found to be small (with the coefficients of the beta function being smaller than one) and thus critical exponents are near-canonical, see, e.g., \cite{Eichhorn:2020sbo}, in particular, see \cite{Pastor-Gutierrez:2022nki} and \cite{Kotlarski:2023mmr} for a discussion of uncertainties. 

\item Non-trivial symmetry identities which are not expected to hold in non-perturbative field theories are approximately satisfied in asymptotically safe gravity-matter theories \cite{Eichhorn:2018akn,Eichhorn:2018ydy,Eichhorn:2018nda}.

\item The gravitational interaction strength is bounded from above, if the generation of free parameters, linked to higher-order matter interactions, is to be avoided \cite{Eichhorn:2016esv,Christiansen:2017gtg,Eichhorn:2017eht,deBrito:2021pyi,Eichhorn:2021qet,Knorr:2022ilz,deBrito:2023myf}.

\end{enumerate}
This near-perturbative nature of asymptotically safe gravity-matter models has two important consequences:

First, it provides us with control over calculations. Concretely, when RG flows are calculated, higher-order interactions are always generated. Neglecting these interactions can lead to unreliable results in non-perturbative settings; but it is expected to be a robust approximation in a near-perturbative regime. We thus base our study on neglecting such higher-order interactions.

Second, a near-perturbative model may allow us to extrapolate our results from Euclidean to Lorentzian signature. Because we use Renormalization Group techniques, we work in a Euclidean setting, where these are much more straightforward to set up (see \cite{Manrique:2011jc,Draper:2020bop,Platania:2020knd,Fehre:2021eob,Banerjee:2022xvi,DAngelo:2022vsh,DAngelo:2023tis,Saueressig:2023tfy} for works in Lorentzian signature). Within a deeply non-perturbative quantum gravity regime, where quantum gravity fluctuations are no longer just fluctuations about a flat background, an analytical continuation from Euclidean to Lorentzian signature is challenging because the configuration spaces of Euclidean and Lorentzian path integrals differ \cite{Knorr:2022mvn} and because a Wick rotation does generically not exist, even at the level of individual spacetime configurations \cite{Baldazzi:2018mtl}. However, in a near-perturbative regime, it is expected that quantum gravity fluctuations are just fluctuations about a flat background about which an analytical continuation, although by no means easy, could be achievable \cite{Bonanno:2021squ,Platania:2022gtt}. We therefore base our Renormalization Group calculations on Euclidean signature in the following.

\section{Abelian Higgs model in asymptotic safety}\label{sec:AbelianHiggs}
We focus on the Abelian Higgs model, which contains an Abelian gauge field $A_{\mu}$ and a charged scalar $\phi$. This is the simplest model which can give rise to cosmic strings \cite{Nielsen:1973cs,Hindmarsh:1994re}. The behavior of the cosmic string network in this model has been analyzed by large-scale numerical simulations \cite{Vincent:1997cx,Moore:2001px,Bevis:2006mj,Bevis:2010gj,Correia:2019bdl}.

We couple it to quantum gravity, in order to understand whether the interplay with quantum gravity is constraining enough to be able to either predict the energy scale relevant for GWs from cosmic strings, or to altogether exclude that cosmic strings are produced. The interplay of asymptotic safety with uncharged scalars has previously been explored in \cite{Narain:2009fy,Narain:2009gb,Zanusso:2009bs,Vacca:2010mj,Eichhorn:2012va,Henz:2013oxa,Percacci:2015wwa,Labus:2015ska,Henz:2016aoh,Eichhorn:2020sbo,Laporte:2021kyp,Laporte:2022ziz}; a dark charged scalar with Higgs portal has been investigated in \cite{Reichert:2019car} and the Higgs scalar in \cite{Shaposhnikov:2009pv, Wetterich:2016uxm, Hamada:2017rvn, Eichhorn:2017ylw, Eichhorn:2018whv,Pawlowski:2018ixd, Wetterich:2019zdo, deBrito:2019umw}.

The effective dynamics for our model, parametrized by scale-dependent couplings, is given by
\begin{widetext}
\begin{equation}
\Gamma_k = \int d^4x\sqrt{g}\Biggl[Z_{\phi}\,g^{\mu\nu}D_{\mu}\phi \left(D_{\nu}\phi\right)^{\dagger} +\bar{m}_k^2 \phi\phi^\dagger + \frac{\bar{\lambda}_{4k}}{4} \left(\phi \phi^{\dagger}\right)^2+\frac{Z_A}{4}g^{\mu\kappa}g^{\nu\lambda}F_{\mu\nu}F_{\kappa\lambda}-\frac{R}{16\pi \, G_N}
\Biggr].
\end{equation}
\end{widetext}

The covariant derivative $D_{\mu}$ contains the gauge connection with gauge coupling $g$, but not the metric connection, because it acts on a spacetime scalar. We make the assumption that the nonminimal coupling of the scalar to the spacetime curvature scalar is negligible and therefore set it to zero in our analysis. Besides two inessential couplings, namely the wave-function renormalizations for the gauge field, $Z_A$, and for the scalar, $Z_{\phi}$, our truncation of the effective dynamics contains the canonically leading terms in the scalar potential, namely mass and quartic interaction, as well as the gravitational coupling $G_N$. The presence of a cosmological constant is unimportant for this setting. 

As it is usual in  searches for asymptotic safety, we introduce dimensionless couplings $(m,\lambda_4,G)$ through 
\be
\bar{m}_k = m\, k\, Z_\phi,\quad  \bar{\lambda}_{4k} = \lambda_4\, Z_\phi^2, \quad G_N = G k^{-2}, 
\ee
where $k$ is the RG-scale, which is a momentum scale. Dimensionless couplings enable us to find scale-symmetry, because they are constant in a scale-symmetric regime. In contrast, dimensionful couplings exhibit scaling, i.e., under changes of the RG scale $k$, they change with $k^{D}$, where $D$ is the mass-dimension of the coupling.

We expand the metric around a flat Euclidean background $g_{\mu\nu} = \delta_{\mu\nu} + \sqrt{32\pi G_N} h_{\mu\nu}$. The resulting gravity-matter vertices give rise to loop diagrams which generate a gravitational contribution to the scale-dependence of the matter couplings. From these diagrams, evaluated with functional RG techniques, we derive the beta functions in order to discover whether there is an asymptotically safe fixed point in the UV, from which a phase of spontaneous symmetry breaking (SSB) can be reached in the IR.\footnote{We do not add finite-temperature effects in our study. These can play a role in the early universe and result in SSB if the low-energy theory is in the symmetry-broken phase and thermal fluctuations generate a positive mass term. We leave the study of such effects to future work.} 

We start by reviewing the beta function for the Abelian gauge coupling, with the quantum-gravity contribution previously calculated in \cite{Daum:2009dn,Folkerts:2011jz,Harst:2011zx,Eichhorn:2017lry,Christiansen:2017cxa,Christiansen:2017gtg}, which is
\be
\beta_g=  - \frac{5}{18\pi} G \,g +\frac{1}{48\pi^2}\frac{g^3}{(1+m^2)^4} + \mathcal{O}(g^4).
\ee
The gravitational contribution comes in at leading order in the gauge coupling and therefore acts akin to a change in the spacetime dimensionality, which would produce a canonical scaling term that is linear in the coupling. The sign of the contribution is negative, such that quantum gravity  antiscreens the Abelian gauge coupling.
Thus, the beta function admits an interacting fixed point at
\be
g_{\ast}= (1+m^2)^2\sqrt{\frac{40\pi}{3}G}.\label{eq:gast}
\ee
This fixed point generates an upper bound for the gauge coupling at the Planck scale, see Fig.~1 in \cite{Eichhorn:2017lry}: any Planck-scale value below $g_{\ast}$ can be reached from the free fixed point, at which the gauge coupling is relevant due to the gravitational effects. The Planck-scale value $g(M_{\rm Planck})= g_{\ast}$ is the unique value that is connected to the interacting fixed point in the UV. Any Planck-scale value above $g_{\ast}$ is not connected to a UV completion (either free or safe). Therefore, the unique asymptotically safe trajectory emanating from $g_{\ast}$ also defines an upper bound for the value of the Abelian gauge coupling at IR scales $g_{\rm IR,max}$. This upper bound is compatible with the observed value since $g_{\rm IR, obs} < g_{\rm IR,max}$ \cite{Eichhorn:2017lry}, see also \cite{Pastor-Gutierrez:2022nki}. 

Next, we turn to the scalar potential and focus on mass and quartic coupling. 
The salient features of the system are already encoded in the leading-order terms of the beta functions at small couplings, see \cite{Eichhorn:2019dhg}. We will, therefore, omit higher-order contributions from $m^2$. For the mass, we have
\be
\beta_{m^2}= - 2m^2- \frac{3}{16\pi^2}g^2 +\frac{53}{18\pi}G\, m^2 - \frac{\lambda_4}{16\pi^2}.
\ee
This beta function admits a fixed point at
\be
m_{\ast}^2 = -\frac{9(3g^2+\lambda_4)}{8\pi \left( 36\pi - 53G \right)}.
\ee
At vanishing quartic coupling, this fixed point lies at negative $m^2$, i.e., in the symmetry-broken regime, unless $G > \frac{36}{53}\pi$ or $g=0$. 

For the quartic coupling, we have the leading-order terms
\be
\beta_{\lambda_4}= \frac{3}{2\pi^2}g^4
- \frac{3}{4\pi^2} g^2 \lambda_4 + \frac{5}{16\pi^2}\lambda_4^2+ \frac{55}{18\pi}G\, \lambda_4.
\ee
Here we already see the key problematic feature from the two terms $\sim g^4$ and $\sim \lambda_4^2$: because both are positive, we do not generically expect real fixed points. The two terms $\sim \lambda_4$ may change this result and, in fact, the beta function has zeros at
\begin{align}
\lambda_{4\, \ast\,, 1/2} = & \frac{2}{45} \left(27g^2- 110\pi G \right. \nonumber \\
 & \left. \pm \sqrt{-1701 g^4-5940\pi g^2\, G +12100\pi^2 G^2} \right).\label{eq:lambdaast}
\end{align}
In order for the fixed-point value to be real, the inequality $-1701 g^4-5940\pi g^2\, G +12100\pi^2 G^2\geq0$ has to be satisfied, which it is, if $ 0 \leq g^2 \leq \frac{110}{189}(-3+\sqrt{30})\pi G $. In particular, there is no interacting fixed point if $G=0$ and the line of fixed-point candidates parameterized by $G$ in Eq.~\ref{eq:gast} lies outside this region. As a result, the fixed-point candidates in Eq.~\ref{eq:lambdaast} lie at complex values, unless one takes $g = 0$, such that $\lambda_{4\, \ast,\, 1}=0$ and $ \lambda_{4\, \ast,\, 2} = -\frac{88\pi}{9}G$. A negative quartic coupling does not automatically imply an instability of the potential; however, it does imply that canonically higher order terms become important to stabilize the potential. We focus on a regime in which the leading-order terms in the potential should suffice; in other words, a perturbative regime. We thus find, in accordance with the analysis in \cite{Eichhorn:2019dhg}, that the Abelian Higgs model with gravity does not feature a nontrivial fixed point with perturbatively stable potential.\\

\begin{figure}[b]
\begin{center}
\includegraphics[width=0.99\linewidth,clip=true, trim=8cm 0cm 10cm 0cm]{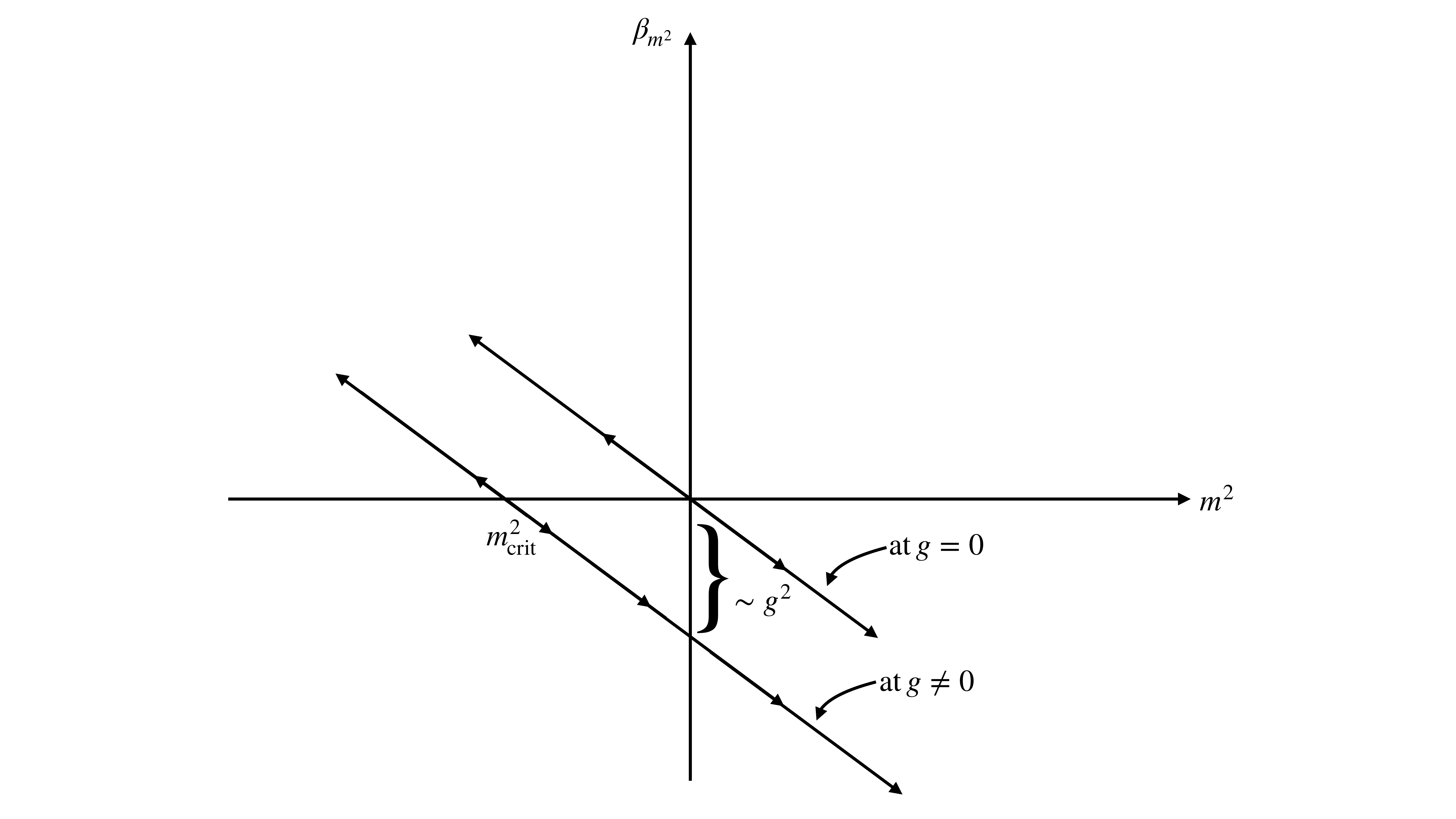}
\end{center}
\caption{\label{fig:illustration_critical_mass}We show the key terms in the beta function for the mass: at $g=0$, the fixed point at $m^2=0$ is infrared repulsive and the mass can decrease toward negative values. At $g\neq0$, the beta function is shifted, so that the mass can only become more negative, if it exceeds the critical value; negative values $m^2>m_{\rm crit}^2$ result in growth of the mass toward positive values.}
\end{figure}

\begin{figure}[t]
\begin{center}
\includegraphics[width=0.99\linewidth]{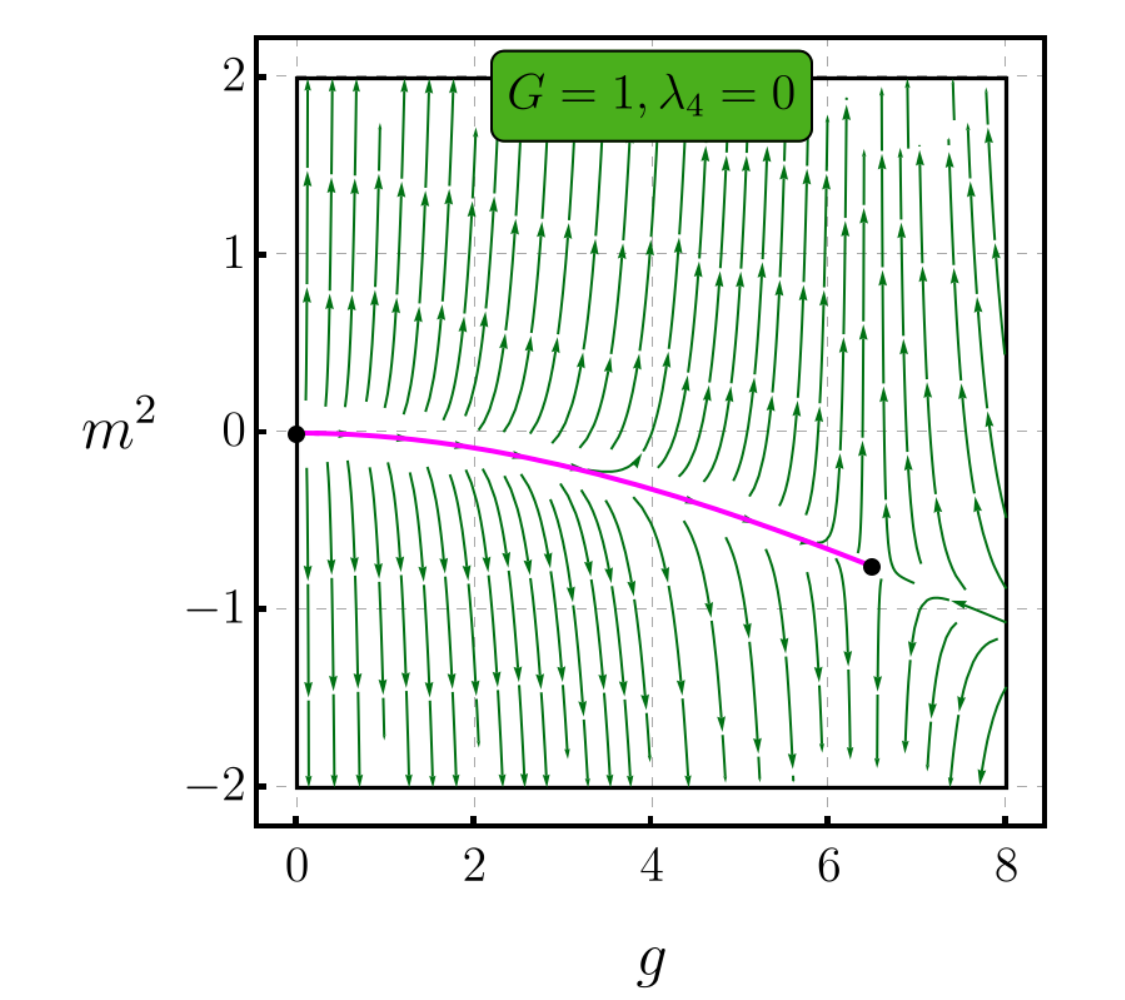}
\end{center}
\caption{\label{fig:Abelian_Higgs_Stream_Plot} We show the phase diagram of the gauge coupling and mass parameter for $G=1$ and $\lambda_{4}=0$, to leading order in $m^2$. The magenta separatrix that connects the free fixed point in the UV to the interacting fixed point in the IR separates the two phases, in which the IR theory features positive $m^2$ (upper region) or negative $m^2$ (lower region). Trajectories beyond $g_*$ (i.e., to the right) are associated with a Landau pole and are not UV complete because quantum gravity fluctuations were not strong enough to antiscreen the gauge coupling in this regime.}
\end{figure}

We therefore turn to the trivial fixed point, at which the gauge coupling is asymptotically free and the potential flat. In this case, the mass parameter remains relevant as long as $G < \frac{36}{53}\pi$, which we assume. In contrast, the quartic coupling is irrelevant. It would thus be predicted to vanish at all transplanckian scales, except for a tiny deviation from zero that is induced, once the mass-parameter departs from zero, see \cite{Eichhorn:2020sbo}. In the following, we do not explore whether or not the resulting potential in the IR is stable. Instead, we only focus on the mass parameter and explore whether, starting from the free fixed point, we can reach a symmetry-broken regime below the Planck scale such that cosmic strings could be generated. To that end, we investigate the beta function for the mass. In the very far UV, the $g^4$-contribution is negligible, because the gauge coupling is asymptotically free. In that regime, $\beta_{m^2}$ is linear in $m^2$, with a negative coefficient. Thus, the mass can deviate from zero toward either positive or negative values and, once it has deviated from zero, continues to grow in magnitude. At the same time, the gauge coupling starts to deviate from zero and increase toward its interacting fixed point value. As long as the mass is more negative than $m_{\ast}$, the mass can counteract the effect of the growing gauge coupling, see Fig.~\ref{fig:illustration_critical_mass} and a phase diagram in Fig.~\ref{fig:Abelian_Higgs_Stream_Plot}.
To estimate the resulting scale of the vacuum-expectation value, which we assume is $v \sim m$, we set $g$ to its fixed-point value, which places the critical value for the mass at
\be
m_{\rm crit}^2 =- \frac{45}{36\pi - 53  G } G.
\ee
To obtain an estimate, we use $G \approx 1$, for which $m_{\rm crit}^2 =-0.75$. This critical value for the dimensionless mass implies that SSB essentially sets in at the Planck scale,\footnote{For a zero-temperature potential with negative mass-squared parameter of the order of the Planck scale, as we obtain here, temperature fluctuations are only expected to restore the symmetry if the temperature is Planckian; thus, the early universe would already be in the symmetry-broken phase.} placing the vacuum expectation value of the field roughly at that scale, and thus too high to be compatible with bounds from the CMB and pulsar timing arrays.
\\

Our analysis up to this point is subject to the caveat of only analyzing the small-coupling regime in $m^2$ and $\lambda_4$. It is not excluded that the large-coupling regime admits symmetry breaking at low enough scales. Going beyond the previous approximations by dropping the perturbative condition in $m^2$ and solving the full system of beta functions, we find no solutions that could change the previous conclusion. In particular, when taking into account mass contributions to the beta function of the quartic coupling, there is  no solution with $m^2_*>0$ and $\lambda_{4*}>0$. Turning to the asymptotically free solution for the gauge coupling, the system is never driven toward a regime of SSB below the Planck scale. We shall, therefore, include extra matter degrees of freedom to search for realizations of a U(1) SSB in the asymptotically safe landscape. 

\section{Extended Abelian Higgs model: the effect of fermions}\label{sec:extAbelianHiggs}

The results in the previous section motivate us to include fermions and extend the Abelian Higgs model. In the presence of fermions, the transplanckian potential can be stable, and fermions can drive the system into a phase of SSB below the Planck scale, just like top quark fluctuations drive the  electroweak sector of the SM toward the electroweak phase transition.

We are not aware of explicit numerical simulations of the cosmic-string network in this extended model, that establish the effect of the fermion. Depending on its mass, it could act as an additional decay channel for cosmic strings, altering the ratio of energy that goes into GWs compared to classical field radiation. We leave such questions to future work, because our main focus is on understanding the asymptotically safe constraints imposed on this model.

\subsection{Yukawa Abelian Higgs model}

The model we explore contains a charged scalar, a charged Dirac fermion and an Abelian gauge field. In our truncation of the dynamics, we neglect non-minimal couplings and higher-order interactions and include the following kinetic terms and interactions 
\begin{align}
\Gamma_k & = \int d^4x \sqrt{g}\, \left[ Z_{\phi}\,D_{\mu}\phi \left(D^{\mu}\phi\right)^{\dagger} +\bar{m}_k^2 \phi\phi^\dagger + \frac{\bar{\lambda}_{4k}}{4} \left(\phi \phi^{\dagger}\right)^2
 \right. \nonumber \\
 & \left. +\frac{Z_A}{4}F_{\mu\nu}F^{\mu\nu} +  i\,Z_\psi \bar{\psi}\gamma_\mu \nabla_\mu \psi
+ i \, \bar{y}_k \left( \phi^* \bar{\psi}^c \psi  + \phi \bar{\psi} \psi^c  \right) \right]  \nonumber \\
& + \Gamma_{k\, \rm EH} + S_{\rm gf} + \Gamma_{k\, \rm SM+BSM}. \label{eq:dynamics}
\end{align}

 As before, the kinetic term for the scalar field contains the gauge coupling $g_\phi$, i.e., $D_\mu = \partial_\mu +  i g_\phi \,Z_A \,A_\mu$. The kinetic term for the fermions contains both the spin connection $\omega_{\mu}^{ab}$ as well as the gauge field, i.e., $\nabla_{\mu} = \partial_{\mu} + \frac{1}{8}[\gamma_a,\gamma_b]\omega_{\mu}^{ab}+ g_e\, Z_A \, A_{\mu}$. For these charged fermions, a Yukawa coupling to the charged scalar can then be written using the charge-conjugated spinor, $\psi^c= -i\, \gamma_2\, \psi^{\ast}$. Moreover, U(1) symmetry requires $g_e=g_\phi/2$. We set $g_\phi = g$.

In this section, we introduce the cosmological constant in the gravitational sector, because it is important for the results in the present section. Thus, our truncation of the gravitational dynamics is given by
\be
\Gamma_{k\, \rm EH} = \frac{-1}{16\pi G_N} \int d^4x\sqrt{g}\left(R- 2 \bar{\Lambda} \right).
\ee
The gauge-fixing term includes a part that gauge-fixes metric fluctuations and a second part that gauge-fixes the Abelian gauge field,
\be
S_{\rm gf }=\int d^4 x \sqrt{\det \bar{g}} \left[  \dfrac{1}{2\alpha'} (\bar{g}^{\mu\nu} \, \bar{D}_\mu A_\nu)^2 + \dfrac{1}{\alpha}\bar{g}_{\mu\nu}\mathcal{F}^\mu\mathcal{F}^\nu \right] ,  \label{eq:gauge}
\ee
where $\bar{D}^\alpha $ is the covariant derivative defined with respect to the background metric $\bar{g}_{\mu\nu}$, which in our case is just the Euclidean metric  $\delta_{\mu\nu}$, and
\begin{align}
	\mathcal{F}^\mu = \left( \bar{g}^{\mu\nu} \bar{D}^\alpha - \dfrac{1+\beta}{4} \bar{g}^{\nu\alpha} \bar{D}^\mu \right)h_{\nu\alpha}.
\end{align}
We work in the Landau-deWitt gauge $\beta=\alpha=0$ and in the Lorenz gauge $\alpha'=0$, and also account for the resulting Faddeev-Popov ghosts.\footnote{For the Abelian gauge sector, the Faddeev-Popov ghosts decouple from the gauge field. They do however not decouple from gravity and thus their fluctuations contribute to the scale-dependence of the gravitational couplings.} Finally, we introduce dimensionless couplings $(m,\lambda_4,y,G,\Lambda)$ through 
$\bar{m}_k = m\, k\, Z_\phi$, $ \bar{\lambda}_{4k} = \lambda_4\, Z_\phi^2$, $ \bar{y}_k = y\,Z_\psi Z_\phi^{1/2}$, $G_N = G k^{-2}$,  and  $\bar{\Lambda} = \Lambda k^{2}$.

We want to establish under which conditions we can induce an SSB of the U(1) symmetry through a vacuum expectation that forms for the scalar at energies below the Planck scale. As we will see, these conditions depend on the gravitational fixed-point values. These in turn depend on the matter content of the theory. To investigate whether there is a set of matter fields that satisfies the conditions, we add extra degrees of freedom on top of our dynamics. These are first all SM fields and second additional BSM fields. None of these fields are coupled to our extended Abelian Higgs model. Accordingly, we neglect self-interactions in these sectors and only consider the minimal coupling of these fields to gravity. This is sufficient to account for their effect on the gravitational fixed-point values.

The corresponding part of the dynamics therefore reads
\begin{align}
\Gamma_{k\, \rm SM+BSM} = \int d^4x\sqrt{g}\,\left[\Bigl(\frac{1}{4}\sum_{I}g^{\mu\kappa}g^{\nu\lambda}{\rm tr}_{I}F_{\mu\nu}F_{\kappa\lambda} \Bigr. \right. & \nonumber \\
 \Bigl.  + \frac{1}{2}g^{\mu\nu}D_{\mu}H^{\dagger}D_{\nu}H + i \sum_{J}\bar{\psi}^J\slashed{\nabla}\psi^{J}\Bigr) & \nonumber\\
 \left. + \Bigl(\frac{1}{4}\sum_{j=1}^{N_V-1} g^{\mu\kappa}g^{\nu\lambda}F^j_{\mu\nu}F^j_{\kappa\lambda} + i \sum_{i=1}^{N_f-1}\bar{\psi}^i\slashed{\nabla}\psi^{i} \Bigr) \right] & . \label{eq:SMplusBSM}
\end{align}
The first line includes a sum over the vectors in the three SM gauge groups, with $I\in$ U(1), SU(2/3), and, in the second line, the Higgs and the SM fermions, with $J$ a superindex that sums over flavors, colors and generations. The covariant derivatives $D_{\mu}$ and $\nabla_{\mu}$ include the gauge covariant derivatives with respect to the corresponding SM gauge groups. For our purposes, the gauge coupling can be neglected. Similarly, the Yukawa couplings in the SM do not play a role in our analysis.

In the third line, we include additional vectors and additional Dirac fermions. Because the extended Abelian Higgs model already contains degrees of freedom beyond the SM, our counting of extra fields goes to $N_V-1$, and $N_f-1$, respectively, i.e., without extra BSM fields in the third line of Eq.~\ref{eq:SMplusBSM}, the system already contains one vector field and one Dirac fermion, in Eq.~\ref{eq:dynamics}.  We assume that the BSM fermions are all charged under the BSM U(1) group, i.e., the covariant derivative $\slashed{\nabla}$ in the third line contains both a spin connection and the Abelian gauge field. The reason for this choice is that an increasing number of charged fermions lowers the fixed-point value for the gauge coupling. This effect will be important, so that we can achieve a \textit{perturbatively} stable potential.

 At the technical level, we add an important result to the existing literature on the gravitational impact on Yukawa couplings: previous work either used a Yukawa coupling between a real scalar and Dirac fermions, or relied on the assertion that the quantum-gravity contribution to all Yukawa couplings is the same, because quantum gravity is blind to internal symmetries. Here, we test this assumption explicitly by calculating the quantum-gravity effect. In the appendixes, we provide details on the calculation.

 Our strategy in this section will be as follows: In a first step, we treat the gravitational couplings as external parameters and search for viable (partial) fixed points in the matter sector as a function of these external parameters. As a result, we obtain a preferred range of values for these external parameters. In a second step, we investigate, whether gravitational fixed-point values fall into this preferred range. This investigation treats the number of fields beyond the SM as variable. We thereby determine a field content beyond the Standard Model, for which the scalar potential is locally stable at the fixed point. Finally, we determine the RG flow starting from this fixed point to show that the mass parameter is relevant and thus, a desired scale of SSB can be accommodated.
 
\subsection{The gauge-Yukawa sector}

As a first step in our analysis, we decouple the scale-dependence of the gauge-Yukawa sector from that of the scalar couplings by setting the scalar mass parameter to zero. Then, $\beta_g$ only depends on $g$ and the gravitational couplings; and $\beta_y$ only depends on $g, y$ and the gravitational couplings. In comparison to the Abelian Higgs model, the beta function of the gauge coupling receives an extra contribution from the BSM Dirac fermions. We use the relation $\beta_g=\frac{g}{2}\eta_A$ \cite{Christiansen:2017gtg,Eichhorn:2017lry}. Therefore, in the perturbative approximation (see Appendix~\ref{app:betafunctions} for full expressions),
\begin{equation}
\beta_g=\left(\frac{5 G}{18 \pi  (1-2 \Lambda )^2} -\frac{5 G}{9 \pi  (1-2 \Lambda)}\right)g+\frac{g^3}{48 \pi ^2 }+N_f\frac{g^3}{48 \pi ^2}. \label{eq:betagauge}
\end{equation}
The fixed-point solutions are $g_*=0$ and 
\begin{equation}
g_*^2 = \frac{40\pi}{3}\frac{(1-4  \Lambda )}{ (1-2 \Lambda )^2} \frac{G}{N_f+1}, 
\label{eq:gsqFP}
\end{equation}
which is positive if $\Lambda<1/4$ and $G>0$. The comparison to the Abelian-Higgs case is straightforward. The interacting fixed point only exists in the presence of gravity. The gauge coupling is relevant around the Gaussian fixed point and irrelevant around the interacting one, producing an upper bound for the gauge coupling at the Planck scale, just as before. The difference between the two cases lies in the contribution of the fermions; as their number increases, the interacting fixed point lies closer and closer to the Gaussian one.

\begin{figure}[!t]
\begin{center}
\includegraphics[width=0.99\linewidth]{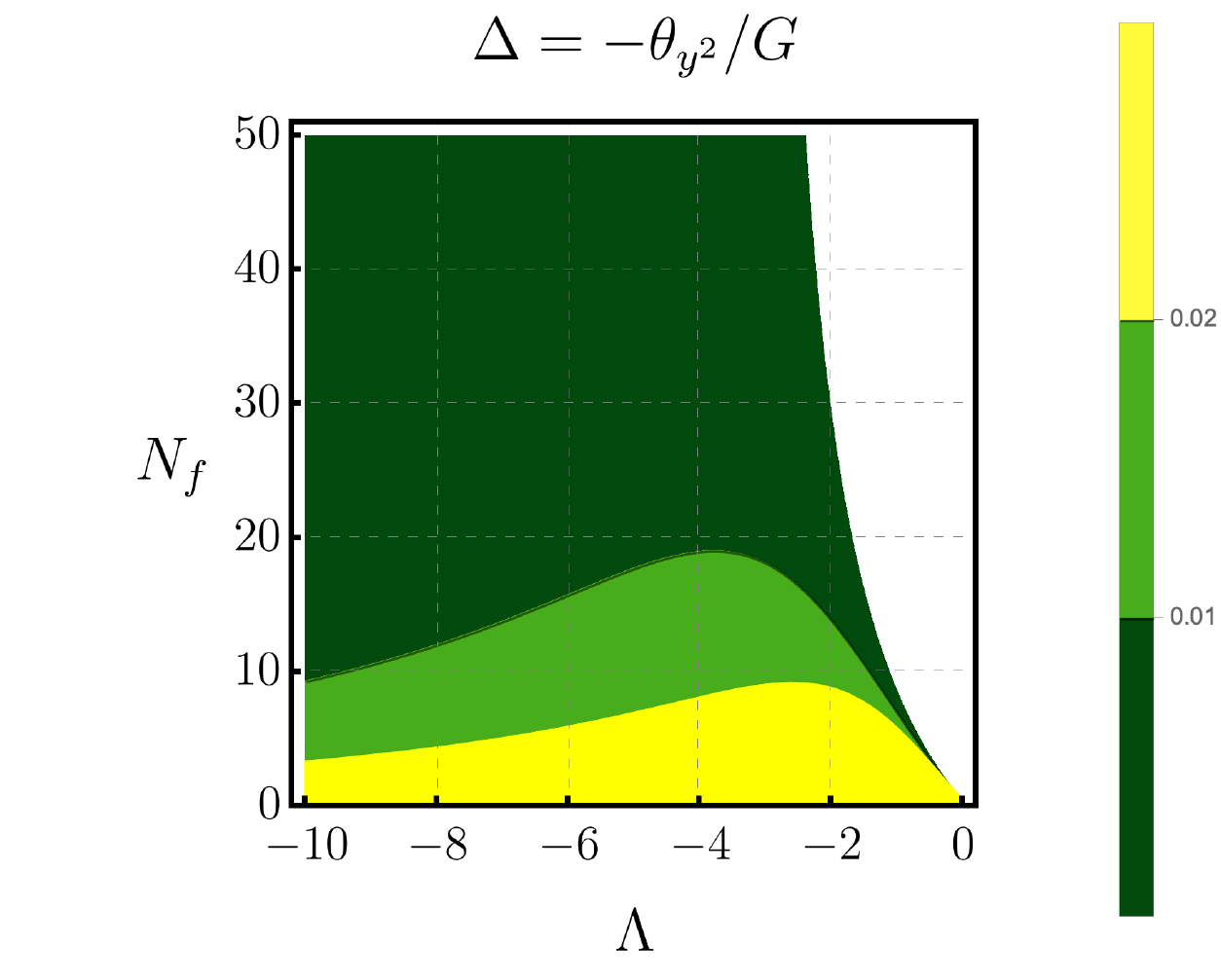}
\end{center}
\caption{\label{fig:crit_exp_yukawa} We show a contour plot for $\Delta>0$ in the $(\Lambda,N_f)$-plane. The line between the dark green and white regions defines $\Lambda_{\star,\rm crit}$, the critical value of $\Lambda$ for a fixed number of  Dirac fermion fields $N_f$. In the colored region, an interacting fixed point for the Yukawa coupling exists and the critical exponents $\theta_{y^2}$, computed around the interacting fixed point, are negative.  We notice that as $N_f$ increases, the absolute value of the critical exponent decreases (for a fixed value of $G$).}
\end{figure}

Next, the beta function for the Yukawa coupling is 
\begin{align}
& \beta_{y^2} = \frac{y^4}{\pi ^2}-\frac{3 g^2 }{16 \pi ^2}\,y^2  \nonumber \\
& +\left(\frac{15 G}{8 \pi  (1-2 \Lambda )^2}- \frac{7 G}{12 \pi  (3-4 \Lambda)}+\frac{7 G}{8 \pi  (3-4 \Lambda )^2} \right) y^2.
\label{eq:betayukawa}
\end{align}
By substituting the value of the interacting fixed point $g_*$ above, we find $y_*^2=0$ and an interacting fixed point
\begin{equation}
y_{\ast}^2 = \pi^2 \Delta\cdot G, \label{eq:ysqfp}
\end{equation}
as long as $\Delta>0$, for
\begin{equation}
\Delta = - \frac{ a + b\, N_f}{ 6\pi (1+N_f) \left(8 \Lambda ^2-10 \Lambda +3\right)^2}, \label{Delta_Yukawa}
\end{equation}
where $a=1016 \Lambda ^3-1577 \Lambda ^2+665 \Lambda -39 $ and $b=56 \Lambda ^3+103 \Lambda ^2-235 \Lambda +96$. The dependence on $N_f$ in $\Delta$ arises from the fixed-point value of the gauge coupling, Eq.~\ref{eq:gsqFP}. As Fig.~\ref{fig:crit_exp_yukawa} shows, this condition is always satisfied if $\Lambda<\Lambda_{\star,\rm crit}$, where $\Lambda_{\star,\rm crit}$ is a function of $N_f$ that becomes increasingly more negative with increasing $N_f$. For instance, for $N_f=1$, $\Lambda_{\star,\rm crit}=-0.10$, while for $N_f=50$, $\Lambda_{\star,\rm crit}=-2.38$.

The critical exponents associated with the Yukawa coupling at the interacting fixed point, $\theta_{y^2}$, and at the Gaussian fixed point, $\theta_{y^2,0}$, are related to each other, $\theta_{y^2,0}= - \theta_{y^2}$, where
\begin{equation}
\theta_{y^2} = - \Delta G\,.
\end{equation}
Therefore, the Yukawa coupling is an irrelevant (relevant) direction around the interacting (Gaussian) fixed point, whenever the interacting fixed points exists. Conversely, if $\Delta$ is negative, the Gaussian fixed point is the only solution and the Yukawa direction is irrelevant, predicting a vanishing Yukawa coupling in the IR. From Fig.~\ref{fig:crit_exp_yukawa}, we see that these critical exponents are much smaller than one. This supports a set of results that indicate that asymptotically safe gravity-matter systems are nearperturbative  and thus characterized by critical exponents close to the canonical scaling dimensions. Further, our choice of truncation, which is based on neglecting canonically higher-order interactions, is thereby a self-consistent approximation of the full system.
 Finally, the perturbative approximation is well suitable as the anomalous dimension of the gauge coupling is zero at the fixed-point solution for $g$, see Eq.~\ref{eq:betagauge}.

In what follows, we focus on the fixed point that is interacting in the gauge and the Yukawa coupling. Because this fixed point is infrared attractive in these two couplings, RG trajectories that start out at one of the other fixed points $(g_{\ast}=0, y_{\ast}\neq 0)$, $(g_{\ast}\neq 0, y_{\ast}=0)$ or $(g_{\ast}=0, y_{\ast}=0)$, are pulled toward the interacting fixed point. Thus, our analysis covers also such crossover trajectories, i.e., trajectories which start out at a (partially) free fixed point, but end up near the interacting fixed point at the Planck scale.

\subsection{The scalar sector}

\begin{figure*}[t]
\begin{center}
\includegraphics[]{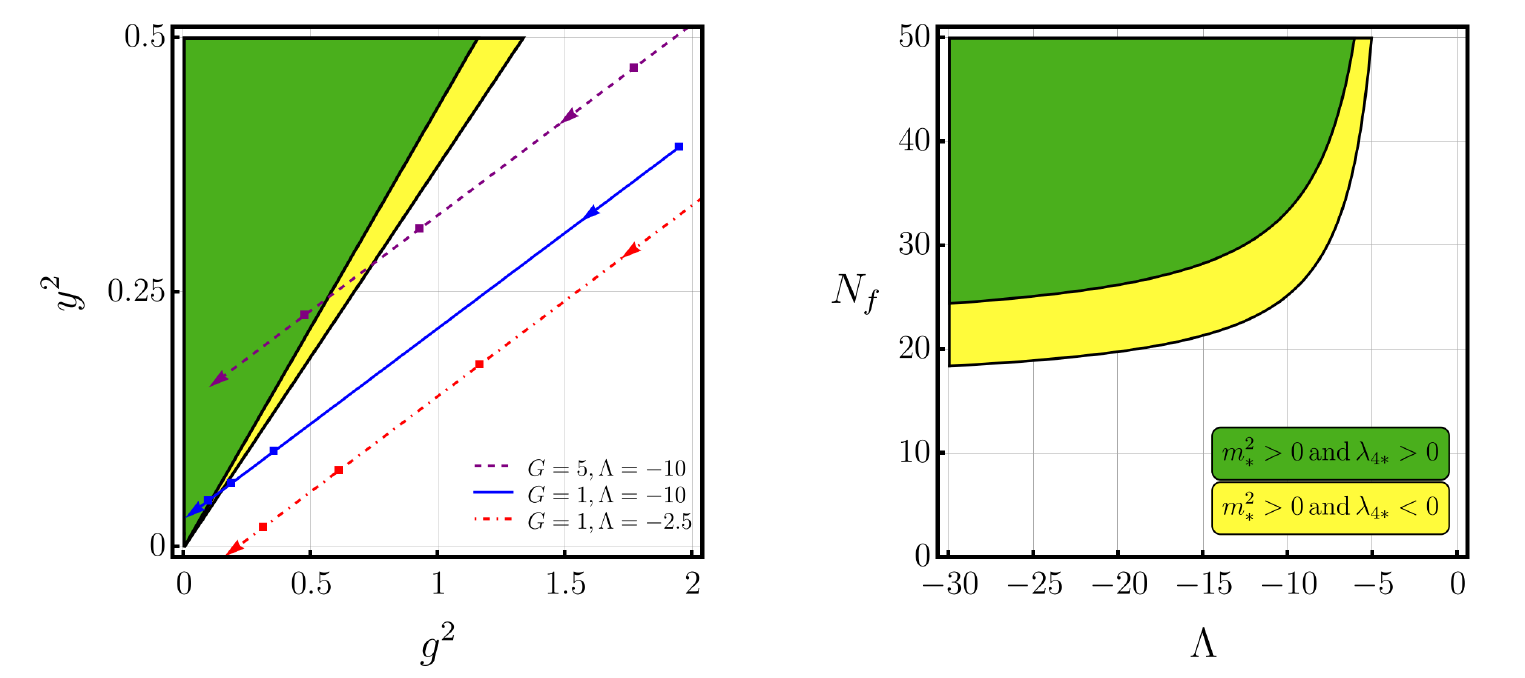}
\end{center}
\caption{\label{fig:criteria} We show in green the region where the two criteria in Eqs.~\ref{eq:criterion1} and \ref{eq:criterion2} are satisfied. In yellow, only the first criterion is satisfied. On the left, we show these regions in the $(g^2,y^2)-$plane and the fixed points $(g^2_*,y^2_*)$ as a function of $N_f$ for $G=1,\Lambda=-2.5$ (red, dot-dashed line), $G=1,\Lambda=-10$ (blue, solid line), and $G=5,\Lambda=-10$ (purple, dashed line). The arrow indicates in which direction  $N_f$ increases. With square markers, we show the fixed points for $N_f=1,10,20,40$ (the fixed point for $N_f=1$ is only shown in the blue, solid line curve). On the right, we show the regions where the criteria are satisfied in the $(\Lambda,N_f)-$plane.}
\end{figure*}

Here we search for UV fixed points with a fixed-point potential that is stable about the origin, i.e., $m_*^2>0$ and $\lambda_{4*}>0$, such that $m^2$ is driven to negative values in the IR, signaling the onset of SSB. The beta functions for $m^2$ and $\lambda_4$ are 
\begin{align}
\beta_{m^2} & = - 2m^2 -\frac{3g^2}{16\pi^2} + \frac{y^2}{2\pi^2} + \left(  - \frac{3g^2}{8\pi^2} + \frac{y^2}{2\pi^2} +\frac{\lambda_4}{8\pi^2} \right) m^2 + \nonumber \\
& + \left( \frac{5 G}{2 \pi  (1-2 \Lambda )^2} + \frac{G}{6 \pi  (3-4 \Lambda)}+\frac{7 G}{2 \pi  (3-4 \Lambda )^2} \right) m^2, \label{eq:betam2} \\ 
\beta_{\lambda_4} & =  \frac{5\lambda_4^2}{16\pi^2} + \frac{3g^4}{2\pi^2} - \frac{8 y^4}{\pi^2} + \left( - \frac{3g^2}{4\pi^2} + \frac{y^2}{\pi^2} \right) \lambda_4 
\nonumber \\
&  + \left( \frac{5 G}{2 \pi  (1-2 \Lambda )^2}+ \frac{G}{3 \pi  (3-4 \Lambda)}+\frac{4 G}{\pi  (3-4 \Lambda )^2} \right)\lambda_4 \nonumber \\
& + \left( \frac{9g^2 }{8\pi^2} -\frac{15 \lambda_4}{16\pi^2} - \frac{50 G }{3 \pi  (3-4 \Lambda)}-\frac{25 G }{\pi  (3-4 \Lambda )^2} \right)m^2\lambda_4. \label{eq:betalambda4}
\end{align}

We demand a perturbatively stable scalar potential and thereby constrain the gauge and Yukawa couplings. We split the analysis in two parts. First, we investigate simplified stability conditions which encode the leading order behavior of these beta functions. We then explore, for which number of fermions $N_f$ the conditions are satisfied. Second, we go beyond leading order and determine fixed-point solutions with perturbatively stable potential numerically.

\begin{figure*}[t]
\begin{center}
\includegraphics[]{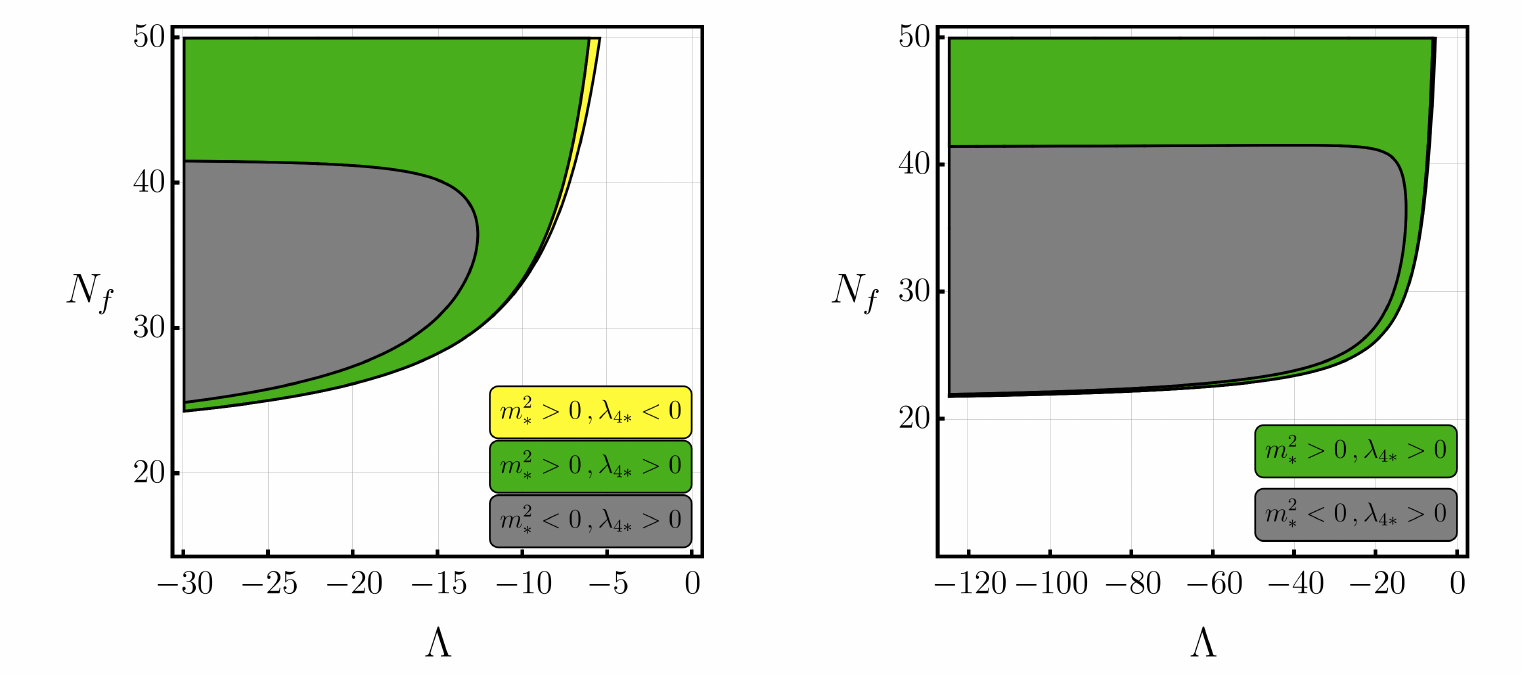}
\end{center}
\caption{\label{fig:nf} We show regions in the $(\Lambda,N_f)$-plane according to the signal of the fixed-point values of the mass and quartic couplings. In both plots, in the green region, the scalar potential is stable around the origin, i.e., both $m^2$ and $\lambda_4$ admit positive fixed-point solutions. In yellow (only visible in the left-hand side plot), only the mass coupling admits a positive fixed point. In gray, only the quartic coupling. In the left plot, we zoom in the region shown in the right plot, where we can see that there is a small window of negative values of $\Lambda$ so that stability is achieved, for $21<N_f<42$. However it is only when $N_f\geq 42$ that we achieve a threshold for $\Lambda$, below which stability is achieved. The value of the threshold increases as the number of fermion fields increases. To produce these plots we can set $G$ to any positive constant value because the regions remain unchanged as we alter $G$ (it changes the value of the fixed points but not their signs).
} 
\end{figure*}

\subsubsection{At leading order}

 To leading order in the mass and quartic coupling, the conditions $m_{\ast}^2>0$ and $\lambda_{4\ast}>0$ become 
\begin{align}
&\text{Criterion 1:} \,-\frac{3g^2}{16\pi^2} + \frac{y^2}{2\pi^2} > 0, \label{eq:criterion1} \\
&\text{Criterion 2:} \,+ \frac{3g^4}{2\pi^2} - \frac{8 y^4}{\pi^2}  < 0.\label{eq:criterion2}
\end{align}
These conditions highlight, that gauge boson fluctuations and fermion fluctuations counteract each other. Fermion fluctuations generate a positive fixed-point value for the mass, because their contribution to $\beta_{m^2}$ is positive. Further, fermionic fluctuations generate a positive fixed-point value for the quartic coupling.

The conditions given by Eqs.~\ref{eq:criterion1} and \ref{eq:criterion2} therefore determine the region in the $(g^2,y^2)-$space where the scalar potential is stable about the origin, depicted on the left-hand side plot in Fig.~\ref{fig:criteria}, where we see that the criterion 2 (in green) is slightly more stringent than the criterion 1 (in the yellow region, only the first criterion is satisfied, while in the green region both are satisfied). Next, we show the fixed-point values for $g$ and $y$ as a function of $N_f$, with $G$ and $\Lambda$ treated as external parameters. A key result of our study is that, unless $N_f$ is very large and the cosmological constant is sufficiently negative, the conditions cannot be fulfilled. 

Next, we substitute the fixed-point solutions $g_*^2$ and $y_*^2$ from Eqs.~\ref{eq:gsqFP} and \ref{eq:ysqfp} in Eqs.~\ref{eq:criterion1} and \ref{eq:criterion2}. Because $g_{\ast}^2$ and $y_{\ast}^2$  are both proportional to $G$, $G$ drops out of the analysis, and we find constraints in the $(N_f, \Lambda)$ plane, cf.~the right-hand side in Fig.~\ref{fig:criteria}. The constraints are only satisfied below a negative threshold value for $\Lambda$, depending on the number of fermions. The larger $N_f$ is, the less negative the threshold becomes. From Eqs.~\ref{eq:criterion1} and \ref{eq:criterion2}, we also find that the criterion 1 is satisfied for $N_f\geq 17$ while the criterion 2 is satisfied for $N_f\geq 22$, no matter how small the cosmological constant is, which is consistent with the flat horizontal directions of the threshold curves in that plot.
\\

We come to the important conclusion that the minimal and most predictive scenario -- an Abelian Higgs model with only one Dirac fermion with a fully interacting fixed point -- is excluded under the assumptions underlying our analysis.\footnote{There is a way to circumvent this result, by choosing the fixed point at which the gauge coupling vanishes and only the Yukawa coupling is interacting. This comes at the cost of introducing an extra free parameter, namely the low-energy value of the gauge coupling, which in this case is only bounded from above.} At least $22$ Dirac fermions are necessary in order to reduce the fixed-point value for $g$ sufficiently in comparison to the fixed-point value for $y$. Next, we drop the simplifying assumptions underlying Eqs.~\ref{eq:criterion1} and \ref{eq:criterion2} to test the robustness of this conclusion. 

\subsubsection{Beyond leading order}

Instead of working to leading order in the mass and quartic coupling in their beta functions, Eqs.~\ref{eq:betam2} and \ref{eq:betalambda4}, we now solve for the zeros of the two beta functions numerically, still under the assumption of negligible mass parameter in the gauge-Yukawa sector, using the interacting fixed-point solutions for $g_*$ and $y_*$.

 We thereby confirm $N_f=22$ as the lower bound on the number of extra Dirac fermions. In addition, we also confirm that the critical value of $\Lambda$ depends on $N_f$ and is less negative for larger $N_f$, cf.~Fig.~\ref{fig:nf}. At the quantitative level, important differences arise.
We find that if $N_f<22$, there are no positive solutions for $m_*^2$ and $\lambda_{4*}$, as is consistent with the simplified criteria. With 22 up to 41 Dirac fermions, we find positive solutions for $\lambda_{4*}$ below a certain threshold of $\Lambda$, as in the simplified case, but now there are only real solutions for $m_*^2$ in a small vertically-oriented strip on the $(\Lambda,G)$-plane, for very large negative values of $\Lambda$. For $N_f=42$, there are solutions for any $\Lambda<-7.2$. The threshold becomes less negative by including even more fermions.
\\

Therefore, dropping the simplified conditions makes the conditions for a perturbatively stable scalar potential  more challenging to meet:  at least 42 Dirac fermions are necessary. This result is quite unappealing for the description of a U(1) SSB below the Planck scale in the asymptotic safety landscape.

\subsection{Yukawa Abelian Higgs model and extra vectors}

We are now finally in a position to include the gravitational fixed-point values, because we have mapped out, where in the space of gravitational fixed-point values necessary conditions for the extended Abelian Higgs model to form cosmic strings are fulfilled.

If we focus on just the Newton coupling, adding extra fermions beyond the Standard Model endangers asymptotic safety. This is because fermions enter $\beta_G$ with the opposite sign of gravitational fluctuations, i.e., they screen the Newton coupling, whereas gravitational fluctuations antiscreen it \cite{Dona:2012am, Dona:2013qba, Meibohm:2015twa, Meibohm:2016mkp, Alkofer:2018fxj, Eichhorn:2018ydy, Eichhorn:2018nda, Wetterich:2019zdo, Daas:2020dyo, Daas:2021abx, Sen:2021ffc}.

When the cosmological constant is added, there is a systematic uncertainty associated to the gravitational fixed-point values under the impact of matter and different approximations may result in somewhat different conclusions, see the discussion in \cite{Eichhorn:2022gku} and references therein. We work in the so-called background-field approximation, in which the background diffeomorphism symmetry of the setting takes center stage.\footnote{Because the region of parameter space that supports a stable potential lies at large negative values of $\Lambda$, fixed-point values in the so-called fluctuation approximation ultimately lead to qualitatively similar conclusions: in that approximation, gravitational fixed-point values do not lie in that region, irrespective of the number of matter fields. In our setting, a sufficiently large number of extra fermions and vectors give access to that region, but also give rise to a highly contrived model. Ultimately, the conclusion appears to hold and cosmic strings from SBB in an Abelian Higgs model are challenging to accommodate in asymptotically safe gravity.} In that setting, the effect of extra fermions can be compensated by adding extra vectors beyond the Standard Model, as one can see from $\beta_G$ according to \cite{Dona:2013qba,
Eichhorn:2020sbo}:
\begin{align}
\beta_G =  2 \,G & - \left( \dfrac{31}{24\pi} + \frac{5 (3-2 \Lambda)}{6 \pi  (1-2 \Lambda )^2} - \frac{ (9-8 \Lambda)}{6 \pi  (3-4 \Lambda)}\right)G^2 \nonumber \\
& -  \left(\frac{4 (N_V-1) - 2 N_f+1}{6 \pi }\right)G^2. \label{eq:beta_g,matter}
\end{align}

Here the effect of SM fields (12 vectors, 4 real scalars and 45 Weyl spinors) and the Abelian Higgs Model (1 vector, 2 reals scalars) is already included and $(N_V-1)$ and $N_f$ denotes the number of vectors and fermions that are additionally included.
For the cosmological constant, we work with the following beta function: 
\begin{align}
\beta_\Lambda = -2 \Lambda &+ \frac{ \left(368 \Lambda ^4-372 \Lambda ^3+320 \Lambda ^2-225 \Lambda +36\right)}{24 \pi  (1-2 \Lambda )^2 (3-4 \Lambda)} G \nonumber \\
  & + \left( \frac{29 + 2 N_f - (N_V-1)}{2\pi} \right) G \nonumber \\
  & -  \left(\frac{4 (N_V-1) - 2 N_f+1}{6 \pi }\right) \Lambda\,G . \label{eq:beta_L,matter}
\end{align}

We find that, as long as $N_f > 42$, there will always be an allowed window for $N_V$ where $G_*>0$ and the scalar potential is stable about the origin. While vector fields compensate the fermionic contributions in $\beta_G$ \cite{Dona:2013qba, Biemans:2017zca, Christiansen:2017cxa, Alkofer:2018fxj, Wetterich:2019zdo, Sen:2021ffc}, they also decrease the fixed-point value of the cosmological constant. As the stability of the scalar potential depends on a negative threshold for $\Lambda$, vector fields can spoil the stability of the scalar potential if the fixed-point value $\Lambda_*$ surpasses that threshold. In particular, for $N_f=42$, a solution $(G_*>0,\Lambda_*<0, m_*^2>0, \lambda_{4*}>0)$ only exists for $21\leq N_V\leq 24$. Therefore in our computation we need at least  41 more Dirac fermions and 20 more vector fields on top of the Yukawa-Abelian-Higgs extension to the SM. 

However, the only reason to introduce this plethora of extra degrees of freedom is to satisfy the criteria of a stable fixed-point potential in the extended Abelian Higgs model. They do not serve any other physical purpose. In particular, the extra vectors cannot be obtained within GUT setting, because the extra scalar fields contained in such a setting drive the fixed-point values out of the regime in which the stability conditions on the scalar potential hold.
\\

We conclude that an Abelian Higgs sector is not easily included in an asymptotically safe gravity-matter setting. This does not rule out cosmic strings. First, one may give up on the highest degree of predictivity and accept additional free parameters. Then, one can choose a fixed point at which the gauge coupling is asymptotically free. From this fixed point, RG trajectories exist, which do not manage to cross over to the interacting fixed point at the Planck scale and at which the gauge coupling is automatically smaller. Then, one only has to satisfy the condition that the Yukawa coupling has an interacting fixed point, which is achievable with fewer extra degrees of freedom, see, e.g., \cite{Eichhorn:2017ylw}.

Second, a genuine GUT setting may contain spontaneously broken U(1) sectors. Because the characterization of asymptotically safe GUTs is quite challenging, if one aims to go beyond more qualitative statements about the degree of predictivity \cite{Eichhorn:2017muy,Eichhorn:2019dhg} and truly characterize the scalar potential \cite{Held:2022hnw}, we refrain from attempting an analysis of cosmic strings from asymptotically safe GUTs in this work.  Additionally, PTA data put stringent bounds on GUTs that predict stable cosmic strings, as the vacuum expectation value $v$ is bounded to be $v \lesssim 4.6 \times 10^{14}$ GeV, which sets an upper bound on the  GUT scale that is lower than typical GUT scales. 

\begin{figure*}[t]
\begin{center}
\includegraphics[]{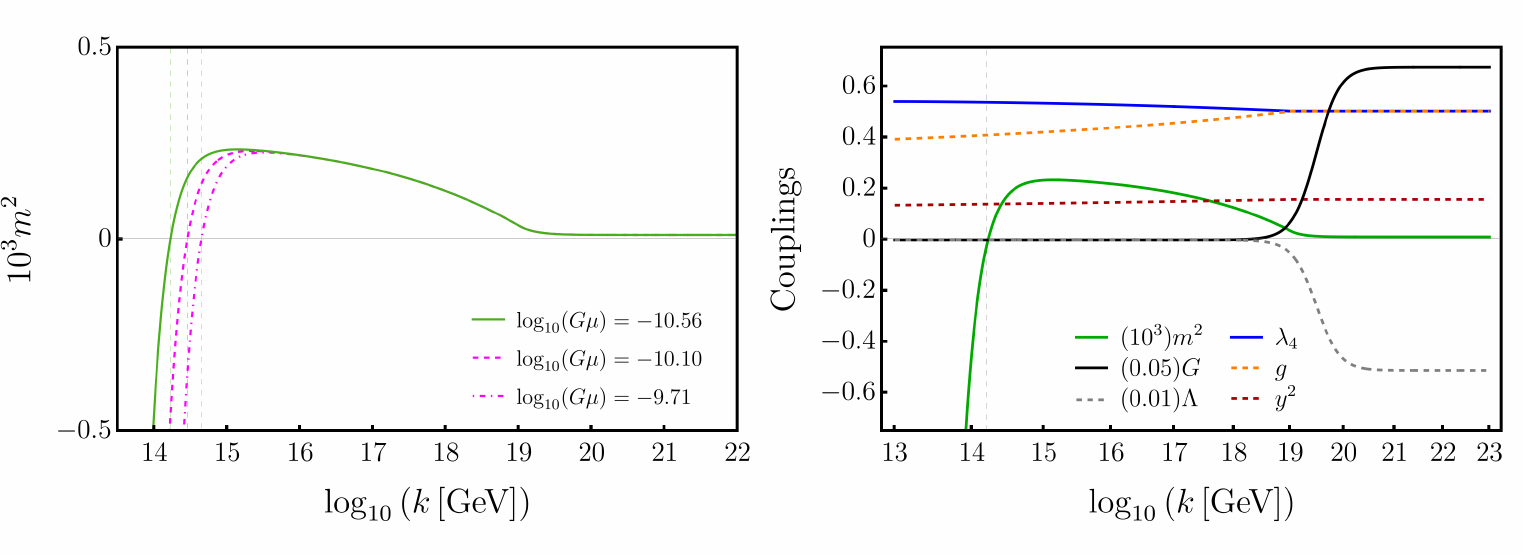}
\end{center}
\caption{\label{fig:flow} On the left, we plot the RG flow of the mass coupling as a function of the RG scale k, for three different choices of string tension following \cite{NANOGrav:2023hvm} (the solid green curve corresponds to the central value of the \textit{monochromatic} stable-cosmic-string model, $\log_{10}(G\mu)=-10.56$, the dashed magenta curve corresponds to the upper bound of the monochromatic stable  cosmic-string-model, $\log_{10}(G\mu)=-10.10$, and the dot-dashed magenta curve corresponds to the upper bound of the \textit{numeric} cosmic-string model, $\log_{10}(G\mu)=-9.71$. On the right, we plot the trajectory of the six (rescaled) couplings $(g,y^2,m^2,\lambda_4,G,\Lambda)$, for that central value of $G\mu$.
}
\end{figure*}

Third, if the additional degrees of freedom we find in our analysis are accepted, one obtains an asymptotically safe extended Abelian Higgs model with a stable fixed-point potential. The fixed-point potential has one free parameter and one dependent quantity in our analysis. The mass parameter is a relevant perturbation of the fixed point and is thus a free parameter of the low-energy theory. The quartic coupling is an irrelevant perturbation and can thus be calculated as a function of the mass parameter. To show that the order of magnitude for the vacuum expectation value $v$, that a cosmic-string interpretation of the NANOGrav data would suggest, can be obtained, we briefly study RG flows from the fixed-point regime into the IR.

\subsection{RG flows}

 Having identified an interacting fixed point with a perturbatively stable potential, we are now in a position to ask whether a regime of SSB, as required for cosmic-string formation, can be reached in the IR. Of the four matter couplings (gauge coupling, Yukawa coupling, scalar mass parameter and scalar quartic coupling), only the scalar mass parameter is a free parameter, because it is the only relevant perturbation of the fixed point. Following our previous discussion, the minimal model requires $N_f=42$ and $N_V=21$.

We obtain the RG trajectory of the six couplings $(g,y^2,m^2,\lambda_4,G,\Lambda)$ solving the set of beta functions numerically in the perturbative approximation. Their beta functions are given by Eqs.~\ref{eq:betagauge}, \ref{eq:betayukawa}, \ref{eq:betam2}, \ref{eq:betalambda4}, \ref{eq:beta_g,matter}, and \ref{eq:beta_L,matter}, respectively. In order to set the scale of SSB, we use the relation between the vacuum expectation value $v$ and the string tension $G\mu$ in Eq.~\ref{eq:vev}.

 The result is shown in Fig.~\ref{fig:flow}. On the left, we show the mass parameter as a function of the RG scale $k$ for three different string tensions $\log_{10}(G\mu) = -10.56$, $\log_{10}(G\mu) = -10.10$, and $\log_{10}(G\mu) = -9.71$, corresponding respectively to the central value of the \textit{monochromatic} stable-cosmic-string model, and the upper bounds for the \textit{monochromatic} and the \textit{numeric} stable-cosmic-string models analyzed in \cite{NANOGrav:2023hvm}. They correspond to the symmetry-breaking scales $v = 1.7 \times 10^{14}$ GeV, $v = 2.8 \times 10^{14}$ GeV, and $v = 4.4 \times 10^{14}$ GeV, respectively, following Eq.~\ref{eq:vev}. The choice for these values is illustrative here to show how a cosmic-string interpretation of the NANOGrav data could be realized in asymptotic safety.\footnote{In field-theoretical simulations of Abelian-Higgs cosmic strings, the scalar-to-gauge mass ratio $\beta = (m^2_\phi/m^2_A) = \lambda_4/ 2g^2$ characterizes the dynamics of the string network: in type-i (type-ii) strings, $\beta < 1$ ($\beta > 1$), there is attractive (repulsive) interaction between parallel vortices \cite{Nielsen:1973cs,Hindmarsh:1994re}, affecting loop formation. The SSB scale $v$  depends on the string tension according to $v =(G\mu / B(\beta) )^{1/2}$, where the critical case $B(1)=1$ gives Eq.~\ref{eq:vev} \cite{Bogomolny:1975de} and, for other values of $\beta$, its precise value must be evaluated with a field-theoretical cosmic string simulation. In general, $B(\beta)$ grows monotonically with $\log\beta$ for type-ii strings \cite{Hill:1987qx,Hindmarsh:1994re}, which is the type of string predominantly investigated in cosmological applications \cite{Vachaspati:2015cma}, effectively lowering the resulting SSB scale.  For works on type-i strings, see, for instance, \cite{Hiramatsu:2013tga,Hindmarsh:2018wkp}.} 

Since the mass is a relevant parameter around its interacting fixed point, we are free to set the SSB scale at these values. In contrast, the quartic coupling is an irrelevant parameter and there is a unique trajectory (prediction) that connects its UV fixed-point value to the IR. We show the trajectory of the mass coupling corresponding to the central value of $G\mu$, together with the other couplings, in the right-hand side in Fig.~\ref{fig:flow}. We identify the scale of SSB with the scale at which the mass parameter crosses zero. This is an order-of-magnitude estimate, and the precise value of $v$ depends on the parameters in the scalar potential. Our work indicates that the nature of the strings is type-ii
(scalar-to-gauge mass ratio $\beta=1.6$ at $v = 1.7 \times 10^{14}$ GeV), which is consistent with Eq.~\ref{eq:vev}, since $\beta$ is sufficiently close to the critical value. To determine these more accurately, the RG flow needs to be continued into the regime of SSB, where additional contributions proportional to the running $v$ occur in the beta functions. Here we do not perform this analysis, because our point is just to show that the free parameter of the setting can indeed be used to choose the SSB scale close to the scale indicated by a cosmic-string interpretation of the PTA data.

\section{Conclusions}\label{sec:conclusions}
We conclude that cosmic strings from a spontaneously broken Abelian gauge theory are not easily accommodated within asymptotically safe quantum gravity. 

First, in the Abelian Higgs model, the scalar potential is not locally stable about the origin in field space. This calls into question whether the Abelian Higgs model, UV completed by the coupling to quantum gravity, is a stable theory. While full potential stability may be achieved by nonperturbative effects, encoded in higher-order operators, a semi-perturbative UV completion without such higher-order operators is ruled out. In particular, in this case the UV theory would be already in the symmetry-broken regime, which likely results in a Planckian value of the vacuum expectation value of the scalar field.

Second, in an extended Abelian Higgs model including a Yukawa coupling to a fermion, we investigate constraints on the gauge and Yukawa coupling which arise from demanding local stability of the potential. We find that those constraints can be fulfilled by the fixed-point values of those couplings at the interacting fixed point, but only in a particular region of the gravitational parameter space. The gravitational fixed-point values only fall into this region if the matter content of the theory is extended: the fixed-point values do not lie in this region when the matter fields corresponding to the extended Abelian Higgs model and the Standard Model are included, but only when a relatively large number of extra degrees of freedom are added. An alternative to introducing so many new fields is to introduce a new free parameter, namely the low-energy value of the gauge coupling. This can be done by starting the RG flow at a partially interacting fixed point, at which the gauge coupling is asymptotically free (and thus relevant).

 Neither of the two options -- new degrees of freedom or new free parameters -- is particularly appealing, because both are only introduced to circumvent the previous, negative result in the Abelian Higgs model.
 \\

In summary, the setting in which an asymptotically safe gravity-matter model gives rise to a spontaneously broken U(1) gauge theory is contrived. The added fields do not serve any phenomenological purpose beyond circumventing the negative result in the simpler model.  The minimal extra field content does not constitute typical matter content of grand unified theories. We thus conclude that a stochastic GW background from cosmic strings is not easily reconciled with asymptotically safe gravity-matter models. If we focus on models which do not contain auxiliary degrees of freedom or auxiliary free parameters, whose only purpose is to circumvent constraints from simpler models, then cosmic strings from an asymptotically safe Abelian Higgs model (with or without a Yukawa sector) are ruled out within the near-perturbative regime of asymptotic safety.

Interestingly, the most recent NANOGrav and EPTA datasets admit the conclusion that stable cosmic strings are not favored by the data \cite{Antoniadis:2023xlr,NANOGrav:2023hvm}. While this cannot be interpreted as a strong hint for asymptotically safe gravity, it is an observation that fits with the results of our study. We stress that a different observational outcome, namely data that favor a stable-cosmic-string interpretation, would have been challenging to accommodate in asymptotically safe gravity and could thus have been interpreted as a hint against asymptotic safety.

In the future, it will be highly interesting to explore whether other BSM settings which give rise to stochastic GW backgrounds, e.g., through first-order phase transitions, can be accommodated in asymptotically safe gravity-matter models \cite{Eichhorn:2020upj}. In this way, PTA data may potentially be used to probe not just particle physics beyond the SM, but even quantum gravity.

\begin{acknowledgments}
This work is supported by a research grant (29405) from VILLUM FONDEN. We thank Gustavo P.~de Brito, Martin Pauly, Marc Schiffer, and Arthur F.~Vieira for useful discussions. RRLdS thanks Kai Schmitz and Tobias Schröder for discussions on cosmic strings and gravitational waves, the group Particle Cosmology Münster at the University of Münster for the warm and extended hospitality, and the NANOGrav Collaboration. JLM thanks Coordenação de Aperfeiçoamento de Pessoal de Nível Superior (CAPES) under Grant No. 88887.478192/2020-00.

\end{acknowledgments}

\appendix

\section{Functional Renormalization Group}
\label{app:FRG}

In this work we search for asymptotic safety in a gravity-matter system using the functional Renormalization Group (FRG). For reviews, see  \cite{Pawlowski:2005xe,Gies:2006wv, Reuter:2012id,Pereira:2019dbn,Bonanno:2020bil,Dupuis:2020fhh,Reichert:2020mja,Pawlowski:2020qer,Saueressig:2023irs,Percacci:2023rbo}. The advantage of this formalism lies in the relatively straightforward computation of the scale-dependence of the couplings of the theory with a (formally) one-loop exact  \textit{flow equation} \cite{Wetterich:1992yh, Morris:1993qb}
\begin{eqnarray}
	k\partial_k \Gamma_k = \dfrac{1}{2}\text{STr}[(\Gamma_k^{(2)}+\textbf{R}_k)^{-1}k \partial_k \textbf{R}_k]. \label{eq:flow}
\end{eqnarray}
Here, the superscript (2)  in $\Gamma_k^{(2)}$ denotes two functional derivatives with respect to the fields, STr is the supertrace ($+1$ for bosonic and $-1$ for fermionic fields), and  $ \textbf{R}_k(p)$ is an \textit{IR regulator} that is introduced in the generating functional $\Gamma_k$ to suppress quantum fluctuations  with momenta $p$ smaller than $k$. This functional  $\Gamma_k$ is the \textit{ flowing action} that interpolates between the bare action (in the limit $k\rightarrow\infty$, i.e., when all quantum corrections are suppressed ) and the full quantum effective action (in the limit $k\rightarrow0$, i.e., when all quantum fluctuations had been taken into account), in a Wilsonian-like approach to the path integral \cite{Wilson:1973jj}. This equation was applied to gravity for the first time by Reuter in his seminal work \cite{Reuter:1996cp}.

The regulator term is diagonal in field space and quadratic in the fields. For each one of the fields, the kernel is
\begin{align}
	\textbf{R}_k(p^2)=
	\left( \Gamma_k^{(2)}(p^2) - \Gamma_k^{(2)}(0)\right)_{h,\phi,A=0}\, r_k(p^2/k^2).
\end{align}
In this work, we chose the Litim regulator \cite{Litim:2001up}, given by $ r_k(y)=\left(\frac{1}{y}-1\right)\theta(1-y)$ for bosons and $r_k(y) = \left(\frac{1}{\sqrt{y}}-1 \right)\theta(1-y)$ for fermions. It respects all global symmetries under which the flowing action is invariant. 

The flowing action $\Gamma_k$ contains all operators that are compatible with the symmetries of the theory.  Since there are an infinite number of such operators, in practice, we truncate $\Gamma_k$ for computational purposes and systematical uncertainties are thus introduced. We base our truncation on the assumption of near-perturbative behavior, see the discussion in the main text. Thus, our truncation includes only canonically relevant and marginal couplings.

\section{Beta functions}
\label{app:betafunctions}

In this section we report the full expressions for the beta functions and anomalous dimensions, which are computed using the flow equation of the Functional Renormalization Group, Eq.~\ref{eq:flow}. The dynamics is given by Eq.~\ref{eq:dynamics}.

\subsection{Anomalous dimensions}
The anomalous dimensions computed from $Z_\phi$, $Z_A$, and $Z_\psi$ are

\begin{widetext}
\begin{align}
\eta_\phi = & - \frac{g_\phi^2 (6-\eta_A)}{32 \pi ^2 \left(1+m^2\right)}-\frac{g_\phi^2 (6-\eta_\phi)}{32 \pi ^2 \left(1+m^2\right)^2}+\frac{y^2(4-\eta_\psi)}{8 \pi ^2} + \frac{G (8-\eta_\phi)}{48 \pi  (3-4 \Lambda) \left(1+m^2\right)^2} \nonumber \\ 
        & + \frac{G (8-\eta_h)}{16 \pi  (3-4 \Lambda )^2 \left(1+m^2\right)}+ \frac{2 G (6-\eta_h) m^2}{\pi  (3-4 \Lambda )^2 \left(1+m^2\right)} -\frac{12 G m^4}{\pi  (3-4 \Lambda )^2 \left(1+m^2\right)^2}, \\
\eta_A = &  \frac{g_\phi^2}{24 \pi ^2 \left(1+m^2\right)^4}+\frac{g_e^2 (4-\eta_\psi) N_f}{24 \pi ^2}-\frac{5 G (8-\eta_A)}{36 \pi  (1-2 \Lambda)}+\frac{5 G (6-\eta_h)}{18 \pi  (1-2 \Lambda )^2}-\frac{5 G (8-\eta_h)}{36 \pi  (1-2 \Lambda )^2},   \\
\eta_\psi = & \frac{ g_e^2 (5-\eta_A)}{80 \pi ^2}-\frac{g_e^2 (4-\eta_\psi)}{64 \pi ^2}+\frac{y^2 (5-\eta_\phi) }{20 \pi ^2 \left(1+m^2\right)^2} - \frac{25 G (6-\eta_h)}{96 \pi  (1-2 \Lambda )^2} + \frac{3 G (6-\eta_\psi)}{40 \pi  (3-4 \Lambda)} + \frac{351 G (7-\eta_h)}{560 \pi  (3-4 \Lambda )^2}-\frac{3 G (6-\eta_h)}{16 \pi  (3-4 \Lambda )^2}.
\end{align}
\end{widetext}

In the main text we employed a perturbative approximation (equivalent to a one-loop perturbative computation), in which we neglect the contributions of anomalous dimensions from regulator insertions. Therefore, we set $\eta_\phi,\eta_A$, and $\eta_\psi$ to zero on the right-hand side in the above equations. The main advantage is to preserve the polynomial form of the beta functions, after substituting the anomalous dimension.

\subsection{Beta functions - matter sector }

For the beta function of the gauge coupling, due to Ward-Takahashi identities \cite{Christiansen:2017gtg,Eichhorn:2017lry}, we use 
\begin{align}
\beta_g = & \frac{g}{2} \eta_A.
\end{align}
The beta function of the Yukawa coupling is 
\begin{widetext}
\begin{align}
\beta_y = & \frac{g_e^2 (6-\eta_A)  y}{32 \pi ^2} + \frac{3 g_e^2 (5 - \eta_\psi ) y}{80 \pi ^2} - \frac{G(6-\eta_\psi ) y}{5 \pi  (3 -4 \Lambda)}+\frac{ G (7-\eta_\psi) y}{56 \pi  (3-4 \Lambda)} + \frac{G (6 -\eta_\psi) m^2 y}{5 \pi  (3-4 \Lambda) \left(1+m^2\right)}-\frac{2G (6-\eta_\phi)  m^2 y}{3 \pi  (3-4 \Lambda) \left(1+m^2\right)^2} \nonumber \\
& +\frac{12G (7-\eta_\phi) m^2  y}{35 \pi  (3-4 \Lambda) \left(1+m^2\right)^2} + \frac{5 G(6-\eta_h)  y}{12 \pi  (1-2 \Lambda )^2} + \frac{ G(6 -\eta_h) y}{2 \pi  (3-4 \Lambda )^2}-\frac{36G (7-\eta_h) y}{35 \pi  (3-4 \Lambda )^2}+\frac{9 G (8-\eta_h) y}{64 \pi  (3-4 \Lambda )^2} \nonumber \\
& -\frac{2 G(6-\eta_h) m^2 y}{\pi  (3-4 \Lambda )^2 \left(1+m^2\right)}  +\frac{36 G(7-\eta_h)  m^2 y}{35 \pi  (3-4 \Lambda )^2 \left(1+m^2\right)}.
\end{align}
\end{widetext}
In the limit of vanishing mass (perturbative 1-loop approximation), we obtain the non-trivial result that the gravitational contribution to $\beta_y$ for our Yukawa term -- invariant under U(1) symmetry -- is the same as it is for a Yukawa term between a real scalar and an uncharged Dirac fermion, computed in \cite{Oda:2015sma,Eichhorn:2016esv}. This result follows the general pattern that gravitational contributions are ``blind" to internal symmetries. 

The beta functions of the mass parameter and quartic coupling are
\begin{widetext}
\begin{align}
\beta_{m^2} = & \, ( \eta_\phi -2 ) \, m^2 - \frac{g_\phi^2 (6-\eta_A)}{32 \pi ^2}-\frac{(6-\eta_\phi) \lambda_4 }{96 \pi ^2 \left(1+m^2\right)^2}+\frac{(5-\eta_\psi) y^2}{10 \pi ^2}  \nonumber \\
& +\frac{5 G (6-\eta_h) m^2}{12 \pi  (1-2 \Lambda )^2} +\frac{ G (6-\eta_h)  m^2}{2 \pi  (3-4 \Lambda )^2}-\frac{2G (6-\eta_h) m^4}{\pi  (3-4 \Lambda )^2 \left(1+m^2\right)}-\frac{2G (6-\eta_\phi)m^4}{3 \pi  (3-4 \Lambda) \left(1+m^2\right)^2}, \\
\beta_{\lambda_4} = & \, 2\eta_\phi \, \lambda_4 + \frac{ (6-\eta_A)g^4}{4 \pi ^2}+\frac{5 (6-\eta_\phi) \lambda_4 ^2}{96 \pi ^2 \left(1+m^2\right)^3}-\frac{8 (5-\eta_\psi) y^4}{5 \pi ^2} +\frac{5 G (6-\eta_h) \lambda_4 }{12 \pi  (1-2 \Lambda )^2}
 +\frac{G (6-\eta_h) \lambda_4 }{2 \pi  (3-4 \Lambda )^2} -\frac{8 G (6-\eta_\phi) \lambda_4  m^2}{3 \pi  (3-4 \Lambda ) \left(1+m^2\right)^2} \nonumber \\
&  -\frac{8 G (6-\eta_h) \lambda_4  m^2}{\pi  (3-4 \Lambda )^2 \left(1+m^2\right)}+\frac{8 G (6-\eta_\phi)  \lambda_4  m^4}{\pi  (3- 4 \Lambda ) \left(1+m^2\right)^3} +\frac{12 G (6-\eta_h) \lambda_4  m^4}{\pi  (3-4 \Lambda )^2 \left(1+m^2\right)^2} +\frac{80 G^2 (6-\eta_h) m^4}{3 (1-2 \Lambda)^3} +\frac{64 G^2 (6-\eta_h) m^4}{(3-4 \Lambda)^3}  
\nonumber \\
& - \frac{256 G^2 (6-\eta_\phi) m^6}{3 (3-4 \Lambda )^2 \left(1+m^2\right)^2} -\frac{512 G^2 (6-\eta_h) m^6}{(3-4 \Lambda )^3 \left(1+m^2\right)} +\frac{1024 G^2 (6-\eta_\phi) m^8}{3 (3-4 \Lambda )^2 \left(1+m^2\right)^3}  +\frac{1024 G^2 (6-\eta_h) m^8}{(3-4 \Lambda)^3 \left(1+m^2\right)^2}.
\end{align}
\end{widetext}

\subsection{Beta functions - gravity sector}

The beta functions of the gravitational parameters $G$ and $\Lambda$ are given by the following expressions from \cite{Eichhorn:2020sbo} are $\beta_G  = 2G + G^2 \, \beta_G^{0} - G^2 M_1(w,s,v)$ and $\beta_\Lambda  = -2\Lambda  + G \, \beta_\Lambda^0  - G M_2(w,s,v) - G \Lambda M_1(w,s,v)$,
where
\begin{align}
\beta_G^{0}= &- \dfrac{31}{24\pi} - \frac{5 (3-2 \Lambda)}{6 \pi  (1-2 \Lambda )^2} + \frac{ (9-8 \Lambda)}{6 \pi  (3-4 \Lambda)},\\
\beta_\Lambda^0= & + \frac{ 368 \Lambda ^4-372 \Lambda ^3+320 \Lambda ^2-225 \Lambda +36}{12 \pi  (1-2 \Lambda )^2 (3-4 \Lambda)},
\end{align}
computed in the Landau-deWitt gauge, whereas $M_1(w,s,v)$ and $M_2(w,s,v)$ are \textit{matter} contributions \cite{Dona:2013qba} given by
\begin{align}
M_1(w,s,v) & = - \frac{s}{6\pi}- \frac{w-4v}{6\pi}, \\
M_2(w,s,v) & = - \frac{s}{4\pi} + \frac{w-v}{2\pi},
\end{align}
where $w$ denotes the number of Weyl fermions, $s$ the number of scalar fields, and $v$ the number of vector fields. For the Standard Model, $w=45$, $s=4$, and $v=12$.
These beta functions are computed in the background approximation.

\bibliography{references.bib}

\end{document}